\documentclass[aps,pra,superscriptaddress,floatfix,groupaddress,english,twocolumn]{revtex4-1}
\usepackage{times,color,amsfonts,amssymb,amsbsy,amsthm,amsmath,graphics,graphicx,hyperref,bm,bbm,makecell,mathtools,dsfont,upgreek,fancyhdr,multirow,newtxmath,subfigure,braket,textgreek}
\DeclarePairedDelimiter{\ceil}{\lceil}{\rceil} 

\usepackage{tcolorbox} 
\hypersetup{colorlinks,linkcolor={blue},citecolor={blue},urlcolor={blue}}
\definecolor{psred}{rgb}{0.8,0.2,0.2}

\begin{document}
\title{Continuous-Variable Measurement-Device-Independent \\ Quantum Key Distribution in Free-Space Channels}
\author{Masoud Ghalaii}
\affiliation{School of Electronic and Electrical Engineering, University of Leeds, Leeds LS2 9JT, United Kingdom}
\affiliation{Department of Computer Science, University of York, York YO10 5GH, United Kingdom}
\author{Stefano Pirandola}
\affiliation{Department of Computer Science, University of York, York YO10 5GH, United Kingdom}

\begin{abstract}
The field of space communications is the realm of communication technologies where diffraction and atmospheric effects, both of which contribute to loss and noise, become overriding. The pertinent questions here are how and at which rate information (secret keys) can be securely transferred (shared) among users under such supposedly severe circumstances. 
In the present work we study continuous-variable (CV) quantum key distribution (QKD) in a measurement-device-independent (MDI) configuration over free-space optical (FSO) links. 
We assess the turbulence regime and provide a composable finite-size key rate analysis of the protocol for FSO links. We study both short-range, horizontal communication links as well as slant paths to, e.g., high-altitude platform station (HAPS) systems.
\end{abstract}

\maketitle

%%%%%%%%%%%%%%%%%%%%%%
\section{Introduction}
Quantum cryptography~\cite{Pirandola:AQCrypt}, one of the oldest quantum technologies, has become a prominent candidate to counteract the challenge from quantum computers~\cite{Gisin:Rev2002}. In particular, quantum key distribution (QKD) has been developing at a rapid pace with the end goal of making distant users able to share a key that must be inscrutable for an eavesdropper to learn about and that therefore can provide highly secure encryption. 
Key challenges for QKD systems include channel loss and noise levels in the communication systems. These are the two main impediments that affect the performance of QKD and its realization, especially over long distances~\cite{Pirandola:PLOB2017}. Until recently, optical fibers have been the main platform to study and experiment most QKD protocols. But their secure distance over long distances is limited, mostly due to the exponential decay of transmissivity in fiber links. In general, two solutions are introduced to conquer this limitation: using quantum repeaters~\cite{Briegel:QRs1998,Dur:QRs1999,Dur:QRs2016,Furrer:QRs2018,Dias:QRs2020,Seshadreesan:QRs2020,Ghalaii:QRs2020} or using free-space and satellite links~\cite{Liao:Nat2017,Ren:Nat2017,Yin:1200kmSat,Bedington:NSat2016,Dequal:njp2021,Mazzarella:QUARC2020,Sidhu:ASpQC}.   

The reach of current terrestrial fibre-based quantum communication systems is limited to only a few hundreds of kilometers~\cite{Zhang:Optica2018}, whereas we seem to stand on the verge of building global quantum communication networks, i.e., quantum internet~\cite{Kimble:QuInternet,Pirandola:QInternet}. 
As a result, recent work has seen a substantial interest in spaceborne QKD and space quantum communications \cite{Sidhu:ASpQC}, aimed at understanding in what way free-space, high-altitude platform station (HAPS) systems, and satellite links may help with current distance limitations, while guaranteeing that quantum safety will be achieved. 
Important steps have been taken, in particular on the limits and security of one-way space quantum communications~\cite{Pirandola:FS2021,Pirandola:Sat2021,Pirandola:CompCVQKD2021}, where it is shown that secret bits can securely be distributed over a turbulent atmosphere, whether weak or strong~\cite{Ghalaii:StTurb2021}. 

At another distinct branch of the QKD science, measurement-device-independent (MDI) QKD~\cite{Braunstein:SideCh2012,Lo:MDI2012} (see also Refs.~\cite{Rubenok:DVMDI2013,Liu:DVMDI2013,daSilva:DVMDI2013} for related experiments) stands as one of the most interesting and well-studied schemes to relax trust assumptions in typical, point-to-point QKD protocols. More precisely, in MDI one does not need to assume that the detection equipment of the legitimate parties, who are going to distribute a secret key between themselves, are trusted. 
This is owing to the fact that a third, allegedly untrusted, party performs the crucial deed of measuring, such that the protocol is immune to all attacks against the measurement modules. What's more, studying MDI protocols over free-space optical (FSO) links is possibly the first step toward investigating space-based quantum repeaters/networks. 
Recently, an experiment implemented discrete-variable MDI, using single photons, over a 19.2~km urban FSO link~\cite{Cao:FSDVMDI2020}. Feasibility studies~\cite{Wang:FSDVMDI2021a,Wang:FSDVMDI2021b} as well as parameter optimization~\cite{Dong:FSDVMDI2021} of space-based discrete-variable MDI QKD with photons were further appeared afterwards.  

Nevertheless, a full security analysis of continuous-variable (CV) MDI protocol that includes parameter estimation and finite-size effects has not yet been presented for the free-space scenario, even though this protocol is known since 2013~\cite{Pirandola:CV2013,Pirandola:CVMDI2015}.
Thus, here we develop the composable security of CV MDI QKD over short FSO links, which are generally affected by diffraction,
atmospheric extinction, turbulence and point errors.
Further we investigate slant paths to mobile devices by studying HAPS systems. 
For all cases we consider the asymmetric configuration where one party is sufficiently close to the MDI station and we compute the composable key rate in the finite size regime.

%%%%%%%%%%%%%%%%%%%%%%%%%%%%
\section{System description}
\begin{figure}[t]   
    \vspace{+.1cm}
    \includegraphics[width=0.5\textwidth-10pt]{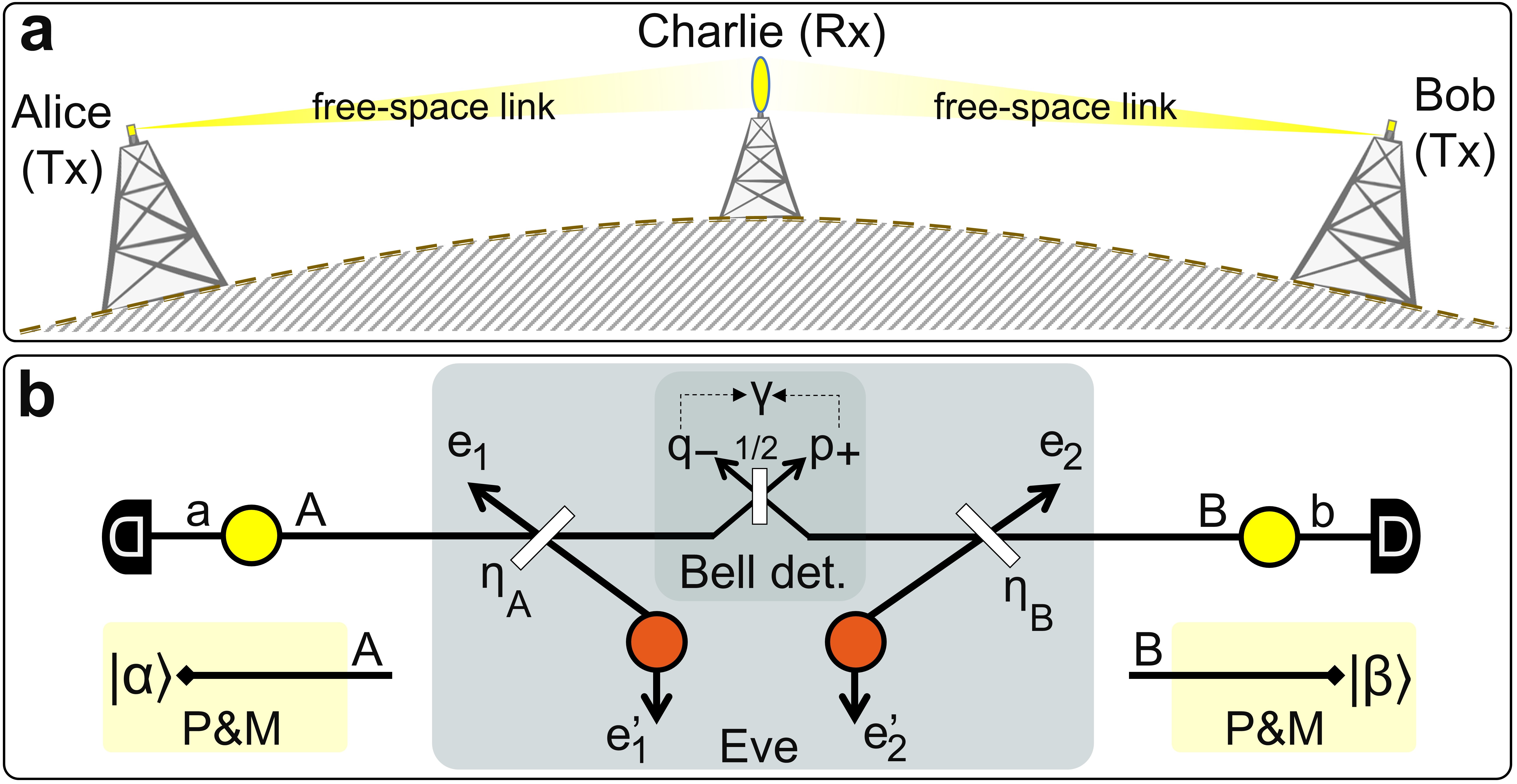}
    \caption{{\bf Schematic of MDI QKD in free space.} 
    {\bf a,} Two parties, Alice and Bob, transmit encoded signals to an untrusted, intermediate party, Charlie, who jointly measures the signals. 
    {\bf b,} The entanglement-based schematic of a continuous-variable protocol with details of the sources and the middle node. Alice and Bob heterodyne one mode of their two-mode squeezed vacuum (TMSV) states, denoted by yellow circles, while subsequently sending the conjugate modes $A$ and $B$ (this is equivalent to the P\&M scheme, where they send a Gaussian-modulated coherent states, e.g., $|\alpha\rangle$ and $|\beta\rangle$). 
    Eve implements an attack by utilizing two TMSV states, denoted by orange circles, and interacting with carrier modes $A$ and $B$. This is modeled via beam splitters of transmissivities $\eta_A$ and $\eta_B$. 
    } 
    \label{fig:FScvMDI}
\end{figure}

Take Alice and Bob to be two terrestrial parties who want to share a quantum-secure key between themselves over FSO links. In an MDI configuration~\cite{Pirandola:CVMDI2015}, they would use two transmitter (Tx) stations and an intermediate receiver station (Charlie, Rx), which is assumed untrusted; see Fig.~\ref{fig:FScvMDI}a. 
They send their modulated coherent-state signals towards the relay Rx, which performs a joint measurement on the received signals and broadcasts the outcome to Alice and Bob. In an {\em asymmetric} MDI setup, the relay is located at an unequal distance from Alice and Bob stations, say it is closer to Alice. 
We assume a Gaussian-modulated protocol, where Alice and Bob choose their quadrature values based on two bivariate Gaussian distributions. 
We also make certain assumptions about the physical FSO channel between the users' and relay's stations. Such assumptions are mainly concerned with the amount of diffraction, pointing error, as well as atmospheric turbulence.
 
In the entanglement-based (EB) scheme of the protocol, as schematically shown in Fig.~\ref{fig:FScvMDI}b, Alice and Bob use two two-mode squeezed vacuum (TMSV) sources that feed Charlie's relay over the FSO links. Charlie is supposed to perform a CV Bell measurement and reports the measurement outcome, $\gamma$, through a public (classical) telecommunication channel.  
Although the altitude of the three stations from sea level can be different, for now we assume that they all are located on top of communication towers with the same height, such that they share a constant-pressure atmospheric turbulence layer. In practice, various effects, including beam-spreading and fading, result in high signal loss which kills the key rate of air-QKD.  

A crucial step in our work is channel modelling. Here we account for diffraction and beam spreading (short/long-term depending on the detectors being fast/slow), background thermal photons, pointing errors and beam wandering. These contribute to have a realistic estimation of the channel loss and channel noise. 
Accordingly, Eve's attack on the FSO links can be modelled by two TMSV states ($e_1e_1'$ and $e_2e_2'$), one mode of each overlapping with Alice and Bob's signals on beam splitters $\eta_A$ and $\eta_B$, respectively. 
We shall do this for single-layer free-space (ground-to-ground) atmospheric paths, where we use specific existing models, such as log-normal. 
We also examine both techniques of measuring CV states, i.e., transmitted local oscillator (TLO) and local local oscillator (LLO).

%%%%%%%%%%%%%%%%%%%%%%
\subsection{Path loss}
The overall optical loss that can occur in a turbulent atmospheric channel can be defined in terms of the multiplication of several types of optical transmissivity
\begin{align}
\label{tot:loss}
\eta(z)= \eta_{\rm eff} \eta_{\rm atm}(z) \eta_\textsc{tb}(z),
\end{align}
where $\eta_{\rm eff}$ is the receiver's efficiency and $\eta_{\rm atm}$ describes the atmospheric loss, which is modelled by the Beer-Lambert equation 
\begin{align}
  \eta_{\rm atm}(z) =\exp\big[-\alpha(\lambda,h) z\big], ~~\alpha(\lambda,h)=\alpha_0(\lambda) e^{-h/6600},
\end{align}
where $h$ is the altitude, in metres, and $\alpha_0(\lambda)$ is the extinction factor at sea level~\cite{Duntley:1948,Bohren:Book}.  
The term $\eta_\textsc{tb}$ is the turbulence-induced transmissivity which, depending on the strength of turbulence, can be computed by several means as we shall discuss in this section.

Let us introduce the dimensionless Rytov variance, which is defined for a plane wave as~\cite{Rytov:1937,Andrews:Book}
\begin{align}
\label{eq:Rytov}
\sigma_\textsc{r}^2(z)=1.23C_n^2k^{7/6}z^{11/6},
\end{align}
where $k=2\pi/\lambda$ is the wavenumber and $C_n^2$ is known as the index-of-refraction structure constant (for a spherical wave the Rytov variance is $0.4\sigma_\textsc{r}^2$). 
For a multiple-layer path, e.g., a slant path form ground to space, the Rytov variance has  a more complex expression. 
For now we restrict our links to be short and within a constant-pressure atmospheric layer, where Eq.~\eqref{eq:Rytov} would suffice.  
It is well accepted that the regime of weak turbulence can be defined by
the condition
\begin{align}
\label{eq:RytovWeak}
\sigma_\textsc{r}^2(z)<1. 
\end{align}
In terms of free-space length, $z$, a more lenient condition,
\begin{align}
\label{eq:z_max}
   z \lesssim z_{\rm max}:=k \min \{4a_{\rm rec}^2,\rho_0^2(z)\},
\end{align}
where $a_{\rm rec}$ is the receiver's aperture radius, 
can be used to describe the strength of the turbulence. Here, 
\begin{align}
    \rho_0(z)=\bigg[0.423 k^2 \int_0^z dz' C_n^2(z') \bigg]^{-3/5}
\end{align} 
is the Fried’s coherence length, which for a constant-pressure atmospheric layer, where $C_n^2$ is constant, reduces to $\rho_0(z)=\big(0.423 k^2 C_n^2 z\big)^{-3/5}$.

We assume a Gaussian beam with initial field spot size $w_0=w(0)$, carrier wavelength $\lambda$ and radius of curvature $F_0$. At distance $z$ of propagation, where a receiver is supposedly placed, free-space diffraction increases the beam's spot size to 
\begin{align}
\label{eq:beamsize}
    w(z)=w_0\sqrt{\Big(1-\frac{z}{F_0}\Big)^2 + \Big(\frac{z}{z_\textsc{r}}\Big)^2},
\end{align}
with $z_\textsc{r}=\pi w_0^2/\lambda$ being the beam's Rayleigh length. Practically, only a fraction of the light can be collected by the receiver, such that the pure diffraction-induced transmissivity is defined as follows
\begin{align}
\label{eq:diff_trans}
    \eta_\textsc{dif}(z)=1-\exp\Bigg[-\frac{2a_{\rm rec}^2}{w^2(z)}\Bigg].
\end{align}

However, the presence of turbulence affects the amount of loss. For the range of distances that we consider in the present paper we do not expect strong turbulence, but wandering of the beam centroid as well as pointing errors can affect the performance. 
On a fast timescale the smaller turbulent eddies deflect the beam. This widens the beam size in Eq.~\eqref{eq:beamsize} to the short-term spot size, $w_\textsc{st}$. This also causes the random Gaussian wandering of the beam centroid with variance $\sigma_\textsc{tb}^2$. 
In addition, pointing errors from jitter and imprecise tracking could cause centroid wandering, such that the centroid quivers with total variance
\begin{align}
\label{eq:TotalWandering}
\sigma^2(z)=\sigma_\textsc{tb}^2(z)+\sigma_\textsc{pe}^2(z).
\end{align}
In other words, the position of the centroid can be taken as a stochastic variable with a Gaussian distribution with variance $\sigma^2$~\cite{Dowling:BeamWandering}. 
The geometric variance of the pointing error at the receiver can be approximated by 
\begin{align}
\sigma_\textsc{pe}^2(z)= \pi \tan^2(\delta/2) z^2, 
\end{align}
where $\delta$, in rad, is the error at the transmitter. For small amounts of $\delta$ one can write $\sigma_\textsc{pe}^2(z) \simeq (\delta z)^2$. We remark that in practice one would collectively estimate the effects of pointing and turbulence on beam wandering~\cite{Avesani:2021}. 

\begin{figure}
\vspace{+.1cm} 
\includegraphics[width=0.5\textwidth-10pt]{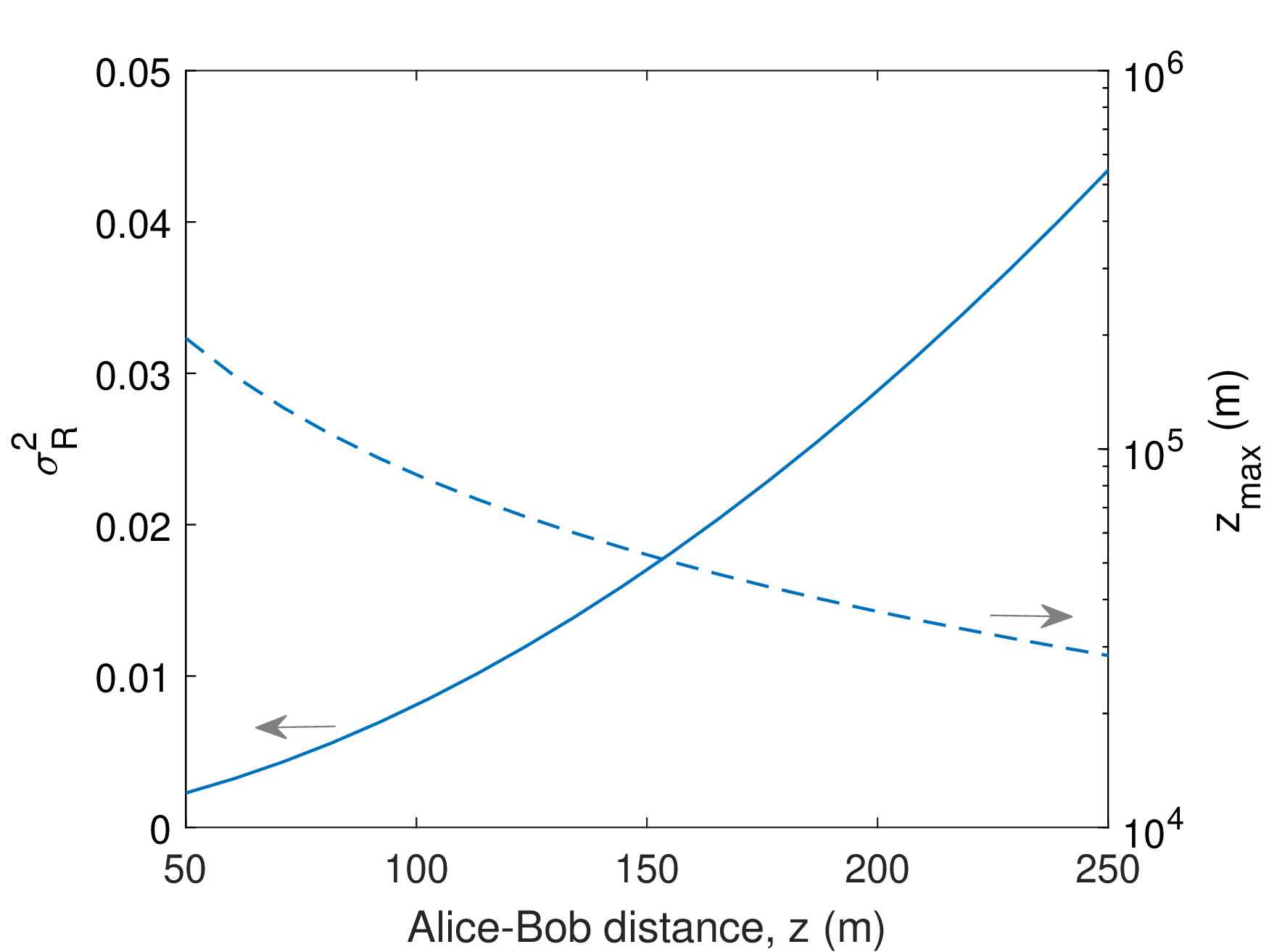}
\caption{{\bf Identifying the regime of turbulence.} 
For the plots we have assumed night-time conditions, with $C_n^2=1.28\times 10^{-14}~\rm m^{-2/3}$, wavelength $\lambda=800~\rm nm$, and aperture size $a_{\rm rec}=20~\rm cm$.
} 
\label{fig:WeakTB}
\end{figure}

Figure~\ref{fig:WeakTB} reveals that the atmospheric turbulence regime we are considering in the present study is indeed weak. Such an atmospheric regime is verified by the help of both conditions given in Eqs.~\eqref{eq:RytovWeak} and \eqref{eq:z_max}. While at all distances considered we have $\sigma_\textsc{r}<1$ we see that $z<z_{\rm max}$ is also verified. This allows us to use Yura's set of equations. 

From Yura's theory~\cite{Yura:1973}, under weak turbulence conditions we have that
\begin{align}
\label{eq:waist_st}
    w_\textsc{st}^2(z)=w^2(z)+\Sigma_\textsc{tb}^2(z), 
\end{align}
where
\begin{align}
    \Sigma_\textsc{tb}^2(z)=2 (1-\phi)^2 \bigg(\frac{\lambda z}{\pi \rho_0(z)}\bigg)^2
\end{align}
accounts for the contribution of turbulence to beam widening. We note that Yura's formulation of weak turbulence regimes requires that $\phi(z):=0.33(\rho_0(z)/w_0)^{1/3} \ll 1$. 
In the present work, we consider a weak satisfaction of this condition ($\phi<0.4$) so that Yura's expansion has to be considered approximate.  
Also, the amount of beam wandering due to turbulence is given by  
\begin{align}
    \sigma_\textsc{tb}^2(z)=\frac{0.1337 \lambda^2 z^2}{w_0^{1/3} \rho_0(z)^{5/3}}.
\end{align}
From the above description, one infers that atmospheric turbulence affects the beam in two ways: the first is by worsening the beam wandering, as described by Eq.~\eqref{eq:TotalWandering}; the second is a diffraction-type effect, as Eq.~\eqref{eq:waist_st} suggests, that results in increasing the beam waist.   

By replacing Eq.~\eqref{eq:waist_st} in the expression for diffraction-induced transmissivity of Eq.~\eqref{eq:diff_trans}, derive
\begin{align}
\label{eq:totloss}
    \eta_\textsc{st}(z)=1-\exp\Bigg[-\frac{2a_{\rm rec}^2}{w_\textsc{st}^2(z)}\Bigg].
\end{align}

However, further modifications are required, e.g., the effect of deflection, which defines wandering of the beam centroid on the receiver's plane following a Gaussian distribution with variance $\sigma^2$. Deflection, with the value $r:=|x_C-x_R|$, where $x_C$ is the location of beam's centroid on the receiver plane and $x_R$ is the aperture center of the receiver, results in an instantaneous transmissivity~\cite{Vasylyev:BeamWandering}
\begin{align}
    \eta_\textsc{st}(z,r)=\eta_\textsc{st}(z) \exp\bigg[-\Big( \frac{r}{r_0}\Big)^\gamma \bigg]. 
\end{align}
Here, we have that 
\begin{align}
\label{eq:gammadef}
    \gamma= & \frac{4\eta_\textsc{st}^{\rm ff}\Lambda_1(\eta_\textsc{st}^{\rm ff})}{1-\Lambda_0(\eta_\textsc{st}^{\rm ff})} \Bigg[\ln \frac{2\eta_\textsc{st}}{1-\Lambda_0(\eta_\textsc{st}^{\rm ff})} \Bigg]^{-1}, \\
    r_0= & a_{\rm rec}\Bigg[\ln \frac{2\eta_\textsc{st}}{1-\Lambda_0(\eta_\textsc{st}^{\rm ff})} \Bigg]^{-1/\gamma},
    \label{eq:r0def}
\end{align}
with $\eta_\textsc{st}^{\rm ff}:=2a_{\rm rec}^2/w_\textsc{st}^2(z)$ being the transmissivity at far field and $\Lambda_n(x)=e^{-2x}I_n(2x)$ ($I_n$ denotes a modified Bessel function of the first kind with order $n$~\cite[Chap.~14]{Arfken}). 

As a result, total transmissivity $\eta$ becomes a function of $r$
\begin{align}
\label{eq:totloss_def}
\eta(z,r)= \eta_{\rm eff} \eta_{\rm atm}(z) \eta_\textsc{st}(z,r),
\end{align}

Consequently, for any physical quantity that is a function of the total transmissivity, such as the key generation rate $K(\eta)$ we have to compute their average 
\begin{align}
\label{eq:AvgQuant}
    \overline{K}(z)=\int_0^{a_{\rm rec}} dr P_{\rm WB}(z,r) K\big(\eta(z,r)\big),
\end{align}
where the expression 
\begin{align}
    P_{\rm WB}(z,r)=\frac{r}{\sigma^2(z)} \exp\Bigg(-\frac{r^2}{2\sigma^2(z)}\Bigg)
\end{align}
is a Weibull distribution for the deflection $r$ and $\sigma^2$~\cite{Pirandola:FS2021}.

%%%%%%%%%%%%%%%%%%%%%%%
\subsection{Path noise}
In general, a receiver sees a total mean number of thermal photons~\cite{Pirandola:FS2021}
\begin{align}
\label{eq:thermalph}
\overline{n}=\eta_{\rm eff}\overline{n}_{\rm bg} + \overline{n}_{\rm ex},
\end{align}
where $\overline{n}_{\rm bg} $ and $\overline{n}_{\rm ex}$ are the number of background thermal photons per mode and extra photons generated within the receiver box, respectively.  
The number $\overline{n}_{\rm bg} $ depends on several factors related to both the sky and the receiver, and is given by
\begin{align}
    \overline{n}_{\rm bg} =\frac{\pi \Gamma_{\rm rec} B_\lambda^{\rm sky}}{\hbar\omega}, 
\end{align}
where $\hbar$ is the reduced Planck constant, $\omega$ is the angular frequency of light, and $B_\lambda^{\rm sky}$ is the brightness of the sky in the range of $10^{-6}-10^{-1}~{\rm Wm^{-2}nm^{-1}sr^{-1}}$ from night to day \cite{Erlong:BackGNoise,Liorni:SatQKD2019}. 
All traces of the receiver are given in
\begin{align}
    \Gamma_{\rm rec}=\Delta\lambda \Delta t \Omega_{\rm fov}a_{\rm rec}^2,
\end{align}
where $\Omega_{\rm fov}$, $\Delta\lambda$, and $\Delta t$ are the angular field of view, spectral filter, and time window of the detector, respectively. The nominal values that we use in the present study are $\Omega_{\rm fov}=10^{-10}~{\rm sr}$, % =10 \mu rad
$\Delta\lambda=0.1~{\rm pm}$, and $\Delta t=10~{\rm ns}$.

We note that the natural interferometric effect of coherent detection, where the signal and local oscillator (LO) pulse overlap, imposes an effective filter of $\Delta\lambda=\lambda^2\Delta\nu/c$, such that assuming $\lambda=800$~nm, a LO of $\Delta t=10$~ns, and a bandwidth $\Delta\nu=50\geq 0.44/\Delta t$~MHz, applies an effective filter of $\Delta\lambda=0.1$~pm.  
This would suppress the background noise $\overline{n}_{\rm bg} $ to the order of $10^{-12}$ ($10^{-7}$) at night (day) time. 
In the asymmetric MDI configuration that we assume in the present study we assume that $\overline{n}_A=\overline{n}_{\rm ex}$ and $\overline{n}_B=\overline{n}$, given by Eq.~\eqref{eq:thermalph}. This is because the distance from Bob to the relay covers almost all the total distance. 
 
Continuous-variable signals, i.e., their quadratures, are measured by means of a homodyne or heterodyne detection, both of which require a reference light, the so-called local oscillator (LO), to perform the detection.  
The LO can be transmitted along with the signal, hence called transmitted LO (TLO), or locally created at the receiver side, hence called local LO (LLO). 
The TLO and LLO schemes add different amounts of noise photons within the receiver. Those generated by LLO, $\overline{n}_\textsc{llo}$, is a linear function of the link transmissivity, while that generated by TLO, $\overline{n}_\textsc{tlo}$, is an inverse function of transmissivity. 
Strictly speaking we have~\cite[Eq.~(62)]{Pirandola:FS2021} 
\begin{align}
\label{eq:noiseLLOTLO}
    \overline{n}_\textsc{llo}=\mathcal{N}+\frac{\pi l_{\rm w} V_A \eta(z)}{C} ~ \text{and} ~ \overline{n}_\textsc{tlo}=\frac{\mathcal{N}}{\eta(z)}, 
\end{align}
where  
\begin{align}
    \mathcal{N}=\frac{\nu_{\rm det}\textsc{NEP}^2 W\Delta t_\textsc{lo} }{2\hbar\omega P_\textsc{lo}}, \notag
\end{align}
with $V_A$ being the modulation variance, $P_\textsc{lo}$ the LO power, $C$ the clock, $l_{\rm w}$ the linewidth, $W$ the detector bandwidth, \textsc{NEP} the noise equivalent power, $\Delta t_\textsc{lo}$ the LO pulse duration, and $\nu_{\rm det}$ the detection noise variance ($\nu_{\rm det}=1$ and $\nu_{\rm det}=2$ for a homodyne and heterodyne detection, respectively). 

\begin{figure}[t]
\vspace{+.1cm}
\includegraphics[width=0.5\textwidth-10pt]{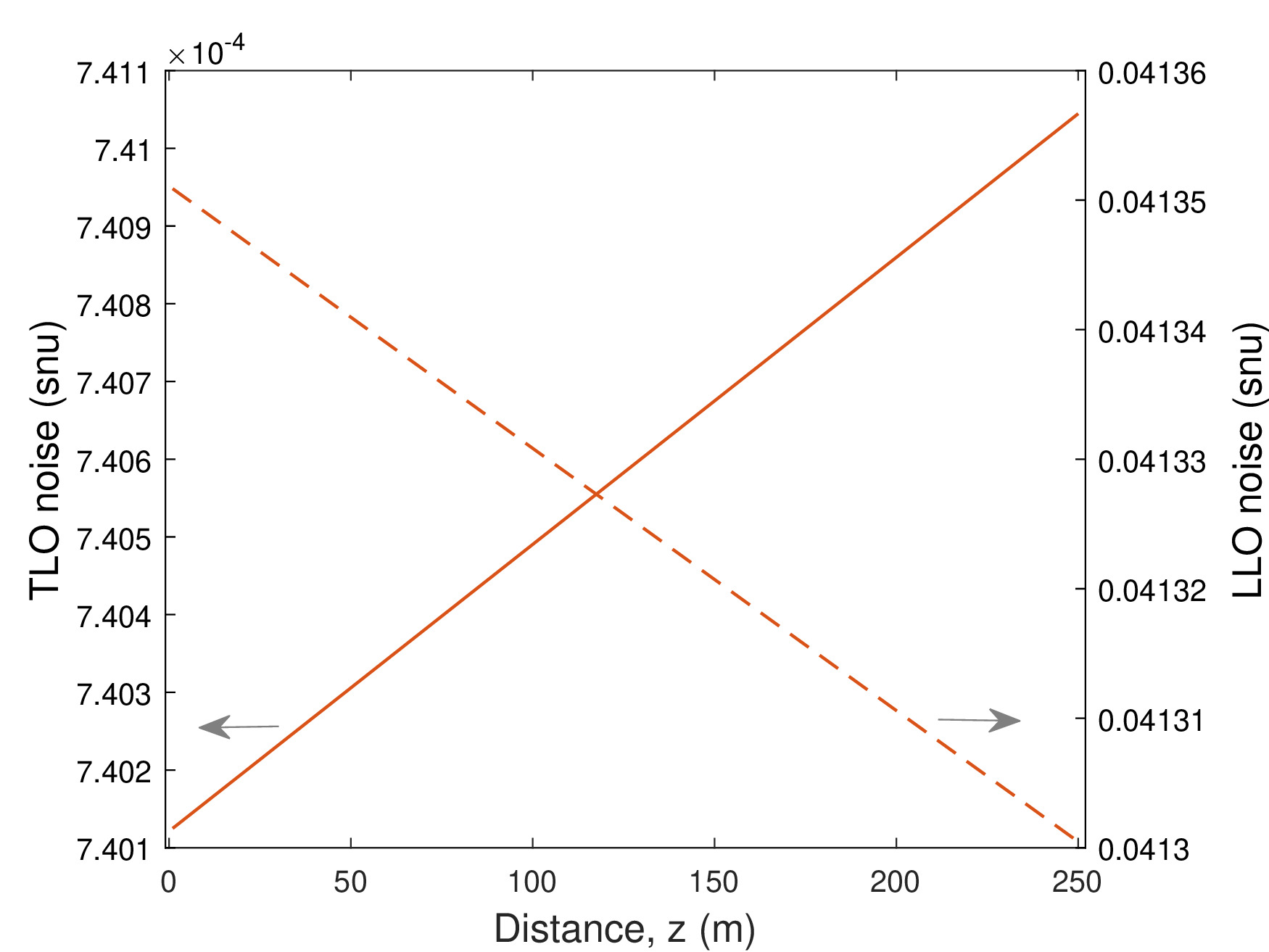}
\caption{{\bf Noise photons generated by a homodyne receiver.} 
We consider night-time, with $C_n^2=1.28\times 10^{-14}~\rm m^{-2/3}$, when a TLO and LLO scheme is used. We have 
   $\lambda=800~\rm nm$ 
   $\textsc{NEP}=6~\rm pW/\sqrt{Hz}$, 
   $W=100~\rm MHz$,
   $\Delta t_\textsc{lo}=10~\rm ns$,
   $P_\textsc{lo}=100~\rm mW$,
   $V_A= 44$,
   $l_{\rm w}=1.6~\rm KHz$,
   $C=5~\rm MHz$, 
   $H_A=H_B=20~\rm m$, $\alpha_0=5\times10^{-6}$, $w_0=10~\rm cm$, $a_{\rm rec}=20~\rm cm$, $\eta_{\rm eff}=0.98$ and $\hbar=1.054\times 10^{-34}~\rm Js$.
} 
\label{fig:RecNoise}
\end{figure}

Figure~\ref{fig:RecNoise} shows the number of extra photons generated at a homodyne receiver. Although at long distances one expects that LLO results in less noise than the TLO~\cite{Ghalaii:StTurb2021}, at short distances TLO introduces about 2 order of magnitudes less noise. 
For the regime of operation we will use in this study we assume the maximum amount of extra noise photons generated at the receiver, that is we assume $\overline{n}_\textsc{llo}=0.04~\rm SNU$.

%%%%%%%%%%%%%%%%%%%%%%%%%%%
\section{Security analysis} 

By using the outcomes of our modelling in the previous sections we can now convey a security analysis by computing achievable key rates for an asymmetric MDI QKD protocol over FSO links.
In the EB representation we assume that Alice and Bob hold two TMSV states with the following covariance matrices (CMs)
\begin{align}
\mathbf{V}_{aA}=
\left(\begin{array}{cc}
\mu_A \mathbf{I} & \sqrt{\mu_A^2-1} \mathbf{Z} \\
\sqrt{\mu_A^2-1} \mathbf{Z} & \mu_A \mathbf{I}
\end{array}\right)
\end{align}
and
\begin{align}
\mathbf{V}_{bB}=
\left(\begin{array}{cc}
\mu_B \mathbf{I} & \sqrt{\mu_B^2-1} \mathbf{Z} \\
\sqrt{\mu_B^2-1} \mathbf{Z} & \mu_B \mathbf{I}
\end{array}\right),
\end{align}
where $\mu_{A(B)}$ defines Alice's (Bob's) TMSV variance. 
By applying heterodyne detection modules to their local modes $a$ and $b$, they project the carrier modes $A$ and $B$ to known Gaussian-modulated coherent states $|\alpha\rangle$ and $|\beta\rangle$, respectively. In other words, Alice and Bob encode the variables $\alpha=(q_A,p_A)$ and $\beta=(q_B,p_B)$ with Gaussian distributions
\begin{align}
G(\alpha)= & \frac{1}{2\pi\sigma_A^2} \exp\Bigg[-\frac{q_A^2+p_A^2}{2\sigma_A^2} \Bigg]
\end{align}
and
\begin{align}
G(\beta)= & \frac{1}{2\pi\sigma_B^2} \exp\Bigg[-\frac{q_B^2+p_B^2}{2\sigma_B^2} \Bigg] 
\end{align}
on the modes $A$ and $B$, such that $\sigma_A^2=\mu_A-1$ and $\sigma_B^2=\mu_B-1$. In the present work we assume equal variances for Alice and Bob, i.e., $\mu_A=\mu_B=\mu$.

On their way through free space, these states experience Eve's attack, which is modelled by means of two beam splitters with transmissivities $\eta_A$ and $\eta_B$. She applies a two-mode attack for each channel by interacting Alice and Bob modes with those of hers that are described by the following CM (see Fig.~\ref{fig:FScvMDI}b)
\begin{align}
\mathbf{V}_{ee'}=
\left(\begin{array}{cc}
\omega_A \mathbf{I} & \mathbf{G} \\
\mathbf{G} & \omega_B \mathbf{I}
\end{array}\right), ~ 
\mathbf{G}=
\left(\begin{array}{cc}
g & 0 \\
0 & g'
\end{array}\right),
\end{align}
where $\omega_A$ and $\omega_B$ quantify Eve’s injected thermal noise while $g$ and $g'$ respect bona fide conditions~\cite{Pirandola:CVMDI2015}. 
The parameters $\omega_A=2\overline{n}_A+\nu_{\rm det}$ and $\omega_B= 2\overline{n}_B+ \nu_{\rm det}$ are total thermal noise variance at Alice-relay and Bob-relay links, respectively, with $\nu_{\rm det}$ being the detection noise variance ($\nu_{\rm det}=1~\rm SNU$ for homodyne and $\nu_{\rm det}=2~\rm SNU$ for heterodyne detection). 
Also, one can argue that the larger $|g|$ and $|g'|$, with $|g|=|g'|$, the stronger the correlation between the modes. 
Then, as the worst-case scenario we can consider an attack with 
\begin{align}
\mathbf{G}_{\rm max}=
\left(\begin{array}{cc}
-g_{\rm max} & 0 \\
0 & g_{\rm max} 
\end{array}\right),
\end{align}
where $g_{\rm max}=\max \{|g|,|g|\}$.

The execution of Charlie's Bell measurement gives the outcome $\gamma=q_C+ip_C$, where $q_C$ and $p_C$ are dependent on
the variables $\alpha$ and $\beta$ 
\begin{align}
    q_C=& - \tau_A q_A + \tau_B q_B + x_N, \\
    p_C=& + \tau_A p_A + \tau_B p_B + p_N, 
\end{align}
where $\tau_{A(B)}= \sqrt{\eta_{\rm eff}\eta_{A(B)}/2}$. 
The variables $x_N$ and $p_N$ are noise variables with variance 
\begin{align}
    \Sigma_N^2=\Xi +\nu_{\rm el} +1,
\end{align}
which includes 1~SNU vacuum noise, electronic noise, $\nu_{\rm el}$, and excess noise  
\begin{align} 
\label{eq:ExNoXi}
    \Xi= & \frac{\eta_{\rm eff}}{2} \big[(1-\eta_A)(\omega_A-1)+(1-\eta_B)(\omega_B-1)\big] \notag \\
    & + \eta_{\rm eff} g_{\rm max} \sqrt{(1-\eta_A)(1-\eta_B)}.
\end{align}

It can be shown that the conditional CM for Alice and Bob is given by~\cite{Papanastasiou:CVMDI}
\begin{align}
\label{CM:ab}
\mathbf{V}_{ab|\gamma}= & 
\left(\begin{array}{cc}
\zeta_a \mathbf{I} & \zeta_c \mathbf{Z} \\
\zeta_c \mathbf{Z}  & \zeta_b \mathbf{I}
\end{array}\right),
\end{align}
where
\begin{align}
    \begin{cases}
    \zeta_a=\mu_A - \frac{\eta_A(\mu_A^2-1)}{\eta_A(\mu_A-1)+\eta_B(\mu_B-1)+2\Sigma_N^2/\eta_{\rm eff}}, \\
    \zeta_b=\mu_B - \frac{\eta_B(\mu_B^2-1)}{\eta_A(\mu_A-1)+\eta_B(\mu_B-1)+2\Sigma_N^2/\eta_{\rm eff}}, \\
    \zeta_c= \frac{\sqrt{\eta_A(\mu_A^2-1) \eta_B (\mu_B^2-1)}}{\eta_A(\mu_A-1)+\eta_B(\mu_B-1)+2\Sigma_N^2/\eta_{\rm eff}}.
    \end{cases}
\end{align}
In addition, a heterodyne detection at Bob's side, with the outcome $\widetilde \beta$, gives the conditional CM at Alice's side
\begin{align}
\label{CM:abcond}
    \mathbf{V}_{a|\gamma \widetilde\beta}= \Big(\zeta_a- \frac{\zeta_c^2}{\zeta_b+1} \Big) \mathbf{I}.
\end{align}

The secret key rate of CV MDI QKD at the asymptotic limit is then given by 
\begin{align}
\label{eq:RateAsym}
    K_\infty(\eta_A,\eta_B,\Xi)=\beta I_{AB}(\eta_A,\eta_B,\Xi) - \chi_E(\eta_A,\eta_B,\Xi), 
\end{align}
where $\beta$ is the reconciliation efficiency, 
\begin{align}
  I_{AB}(\eta_A,\eta_B,\Xi)=  \frac{1}{2}\log_2\frac{1+\det \mathbf{V}_{a|\gamma}+ {\rm tr} \mathbf{V}_{a|\gamma} }{1+\det \mathbf{V}_{a|\gamma \widetilde\beta}+ {\rm tr} \mathbf{V}_{a|\gamma \widetilde\beta}}
\end{align}
is mutual information, 
and
\begin{align}
  \chi_E(\eta_A,\eta_B,\Xi)= g(\nu_+)+g(\nu_-)+g(\nu_{\rm c})
\end{align}
is Holevo information, with $\nu_\pm$ being eigenvalues of the CM $\mathbf{V}_{ab|\gamma}$ and $\nu_{\rm c}$ being the eigenvalue of the conditional CM  $\mathbf{V}_{b|\gamma \widetilde{\alpha}}$, and we define
\begin{align}
    g(x)=\frac{x+1}{2}\log_2\frac{x+1}{2} - \frac{x-1}{2}\log_2\frac{x-1}{2}.
\end{align}

The stochastic nature of free-space channels causes fluctuations that result in free-space fading. Hence, the transmissivities, as well as the level of noise, become unstable and vary with time over certain time scales, such that the probability distribution for the deflected transmissivity is~\cite{Pirandola:FS2021} 
\begin{align}
    P_0(\tau)=\frac{r_0^2}{\gamma \sigma^2\tau} \Big(\ln \frac{\eta}{\tau}\Big)^{2/\gamma-1} \exp\bigg[-\frac{r_0^2}{2\sigma^2} \Big(\ln \frac{\eta}{\tau}\Big)^{2/\gamma} \bigg], 
\end{align}
where $\gamma$ and $r_0$ are given in Eqs.~\eqref{eq:gammadef} and \eqref{eq:r0def}, respectively, and the mean-value of deflection is assumed to be zero. 
Thus, estimated parameters and the key rate would take different values than that given in Eq.~\eqref{eq:RateAsym}. 
The details of such an issue was introduced in Refs.~\cite{Pirandola:FS2021,Pirandola:CompCVQKD2021} for one-way CV QKD protocols. Also, as a possible solution, the pilot pulses were introduced. 
In the following, we explain how one can use the pilot solution in the case of free-space CV MDI QKD protocols. 

Pilot pulses are relatively intense pulses that help to track and measure/estimate the instantaneous transmissivity. The pilots are weak enough to be measured via LO signals but much brighter than quantum signals to provide a good estimate of the transmissivity. In fact, they help to collect signals within a lattice of suitable time-bins with almost equal transmissivity. 
Therefore, in a free-space scenario, apart from $m_\textsc{pe}$ samples that are sacrificed for parameter estimation (PE), $m_\textsc{pl}$ of the signals are energetic pilot signals that are used to estimate the instantaneous transmissivity, such that $N=n+(m_\textsc{pe}+m_\textsc{pl})$, where $n$ will be consumed for building the raw key.

For our MDI setup, let us assume that both Alice and Bob send coherent-state pilots $|\overline{n}_\textsc{pl} \rangle$ towards the relay, which treats pilots as normal quantum signals, i.e., it outcomes $\gamma_\textsc{pl}$. This would allow Alice and Bob to build the estimators for the instantaneous transmissivities $\tau_{A(B)}= \sqrt{\eta_{\rm eff}\eta_{A(B)}/2}$. 
In a fading interval $[\tau,\tau+\delta \tau]$, a fraction of the pilots $p_\delta m_\textsc{pl}$, where 
\begin{align}
    p_\delta:=\int_\tau^{\tau+\delta \tau} P_0(\tau) d\tau,  
\end{align}
can be used for estimating $\tau_A$ and $\tau_B$.  
From the pilots, the number of $p_\delta m_\textsc{pl} \nu_{\rm det}$ outcome pairs $(q_{C,i},p_{C,i})$ of the relay, i.e., 
\begin{align}
    q_{C,i}=& - \tau_A q_{A,i} + \tau_B q_{B,i} + x_{N,i}, \\
    p_{C,i}=& + \tau_A p_{A,i} + \tau_B p_{B,i} + p_{N,i}, 
\end{align}
where $q_{A,i}=p_{A,i}=q_{B,i}=p_{B,i}=\sqrt{2\overline{n}_\textsc{pl}}$, can be derived.
Alice and Bob can then build the estimators 
\begin{align}
    \hat{T}_{A,\textsc{pl}}:= & \frac{1}{p_\delta m_\textsc{pl} \nu_{\rm det}}\sum_i \frac{-q_{C,i}+p_{C,i}}{2\sqrt{2\overline{n}_\textsc{pl}}},   \\
    \hat{T}_{B,\textsc{pl}}:= & \frac{1}{p_\delta m_\textsc{pl} \nu_{\rm det}}\sum_i \frac{q_{C,i}+p_{C,i}}{2\sqrt{2\overline{n}_\textsc{pl}}},
\end{align}
with mean $\tau_A$ and $\tau_B$, respectively, and variance $\sigma_N^2/(8 p_\delta m_\textsc{pl} \nu_{\rm det} \overline{n}_\textsc{pl})$. It can be argued that real-time tracking of the transmissivities is possible with negligible error for a sufficiently large $\overline{n}_\textsc{pl}$, even if $m_\textsc{pl}$ is small. 

While it is possible to introduce postselection intervals $[\tau_{A,\min},\tau_{A,\max}]$ and $[\tau_{B,\min},\tau_{B,\max}]$, the parties can choose the minimum achievable values $\tau_{A,\min}$ and $\tau_{B,\min}$ to wipe out the fading and build a stable link. 
Following Refs.~\cite{Pirandola:FS2021,Pirandola:CompCVQKD2021}, we take $\tau_{B,\min}=f_{\rm th} \eta_B$, where $f_{\rm th}$ is a fixed postselection threshold. At the same time, for a very asymmetric MDI protocol, one can assume that $\tau_{A,\min}=\eta_A$.  
These values can also modify associated noise values given in Eq.~\eqref{eq:noiseLLOTLO} as well as the excess noise given in Eq.~\eqref{eq:ExNoXi}. Therefore, the secret key rate at the asymptotic limit in Eq.~\eqref{eq:RateAsym} will be given by $K_\infty(\eta_{A,\min},\eta_{B,\min},\Xi_{\max})$.

To deliver a more rigorous account of the key rate analysis, in the following we compute the composable finite-size key rate analysis by also presenting the PE step. 
We assume that Alice and Bob use $m_\textsc{pe}$ samples for PE. 
Accepting an error $\epsilon_\textsc{pe}$, which is the error probability associated with each estimator, one can provide the following worst-case scenario values for the transmissivities and the excess noise (here for convenience we drop the `$\min$' and `$\max$' subscripts from the transmissivities and the noise)
\begin{align}
\label{eq:eta_Atilde}
    \widetilde \eta_A= & \eta_A-w\sqrt{\sigma_{\eta_A}^2}, \\
    \widetilde \eta_B= & \eta_B-w\sqrt{\sigma_{\eta_B}^2}, \\
    \widetilde \Xi= & \Xi+w\sqrt{\sigma_N^2},
\end{align}
where $w=\sqrt{2}\text{erf}^{-1}(1-\epsilon_\textsc{pe})$, $\Xi=\Sigma_N^2-\nu_{\rm el}-1$, and
\begin{align}
    \sigma_{\eta_A}^2 \simeq & \frac{16 \eta_A}{m_\textsc{pe}} \Bigg[\eta_A + \frac{\eta_B \sigma_B^2}{2\sigma_A^2} \Bigg] \Bigg\{1+\frac{\Sigma_N^2/\eta_{\rm eff}}{\eta_A\sigma_A^2+\eta_B\sigma_B^2/2} \Bigg\}, \\
    \sigma_{\eta_B}^2 \simeq & \frac{16 \eta_B}{m_\textsc{pe}} \Bigg[\eta_B + \frac{\eta_A \sigma_A^2}{2\sigma_B^2} \Bigg] \Bigg\{1+\frac{\Sigma_N^2/\eta_{\rm eff}}{\eta_B\sigma_B^2+\eta_A\sigma_A^2/2} \Bigg\}, \\
    \sigma_N^2 \simeq & \frac{2(\Sigma_N^2)^2}{m_\textsc{pe}}.
\end{align}

Thus, the worst-case, minimum secret key rate based on the PE scheme is given by 
\begin{align}
\label{eq:RatePE}
    K_\textsc{pe}(\widetilde\eta_A,\widetilde\eta_B,\widetilde\Xi)=\beta I_{AB}(\widetilde\eta_A,\widetilde\eta_B,\widetilde\Xi) - \chi_E(\widetilde\eta_A,\widetilde\eta_B,\widetilde\Xi).
\end{align}

What's more, the key rate must be composably secure~\cite{Pirandola:AQCrypt}, 
including imperfections in the data processing~\cite{Papanastasiou:2022}. 
Assuming that the free-space link is used $N$ times, the composable finite-size is given by~\cite{Pirandola:FS2021}
\begin{align}
\label{eq:RateComp}
    K(z,r)=\frac{np_\textsc{ec}}{N} \Bigg(K_\textsc{pe} (\widetilde\eta_A,\widetilde\eta_B,\widetilde\Xi)-\frac{\Delta_\textsc{aep}}{\sqrt{n}}+\frac{\Theta}{n} \Bigg), 
\end{align}
where~\cite{Pirandola:FS2021,Pirandola:CompCVQKD2021}
\begin{align}
    \Delta_\textsc{aep}:= & 4\log_2(\sqrt{d}+2) \sqrt{\log_2(18 p_\textsc{ec}^{-2} \epsilon_\textsc{s}^{-4} )} , \\
    \Theta:= & \log_2\Big[p_\textsc{ec}(1-\frac{\epsilon_\textsc{s}^2}{3})\Big] + 2\log_2(\sqrt{2}\epsilon_\textsc{h}).
\end{align}
Equation~\eqref{eq:RateComp} gives the rate for a protocol with overall security parameter $\epsilon= \epsilon_\textsc{c} + \epsilon_\textsc{s} + \epsilon_\textsc{h} + 3 p_\textsc{ec}\epsilon_\textsc{pe}$. Assuming reverse reconciliation, the hash comparison step of the finite-key analysis requires Bob to send $\ceil[\big]{\log_2(1-\epsilon_\textsc{c})}$ bits to Alice for proper values of $\epsilon_\textsc{c}$ (called $\epsilon_\textsc{c}$-correctness) and bounds the probability that Alice's and Bob's sequences differ even if their hashes match. Also, the $\epsilon_\textsc{h}$ and $\epsilon_\textsc{s}$ describe errors that occur during the hashing and the smoothing stages, respectively. It is also convenient to define the frame error rate $\textsc{FER}=1-p_\textsc{ec}$. Further, it is assumed that by using an analog-to-digital conversion each continuous-variable symbol is encoded with $d$ bits of precision. 
We remark that since the transmissivities are dependent on the deflection parameter $r$, such that the rate in Eq.~\eqref{eq:RateComp} is a function of $r$, one needs to use the integral in Eq.~\eqref{eq:AvgQuant} to compute an average rate. 

\begin{figure}[t]
\vspace{+.1cm}
\includegraphics[width=0.5\textwidth-10pt]{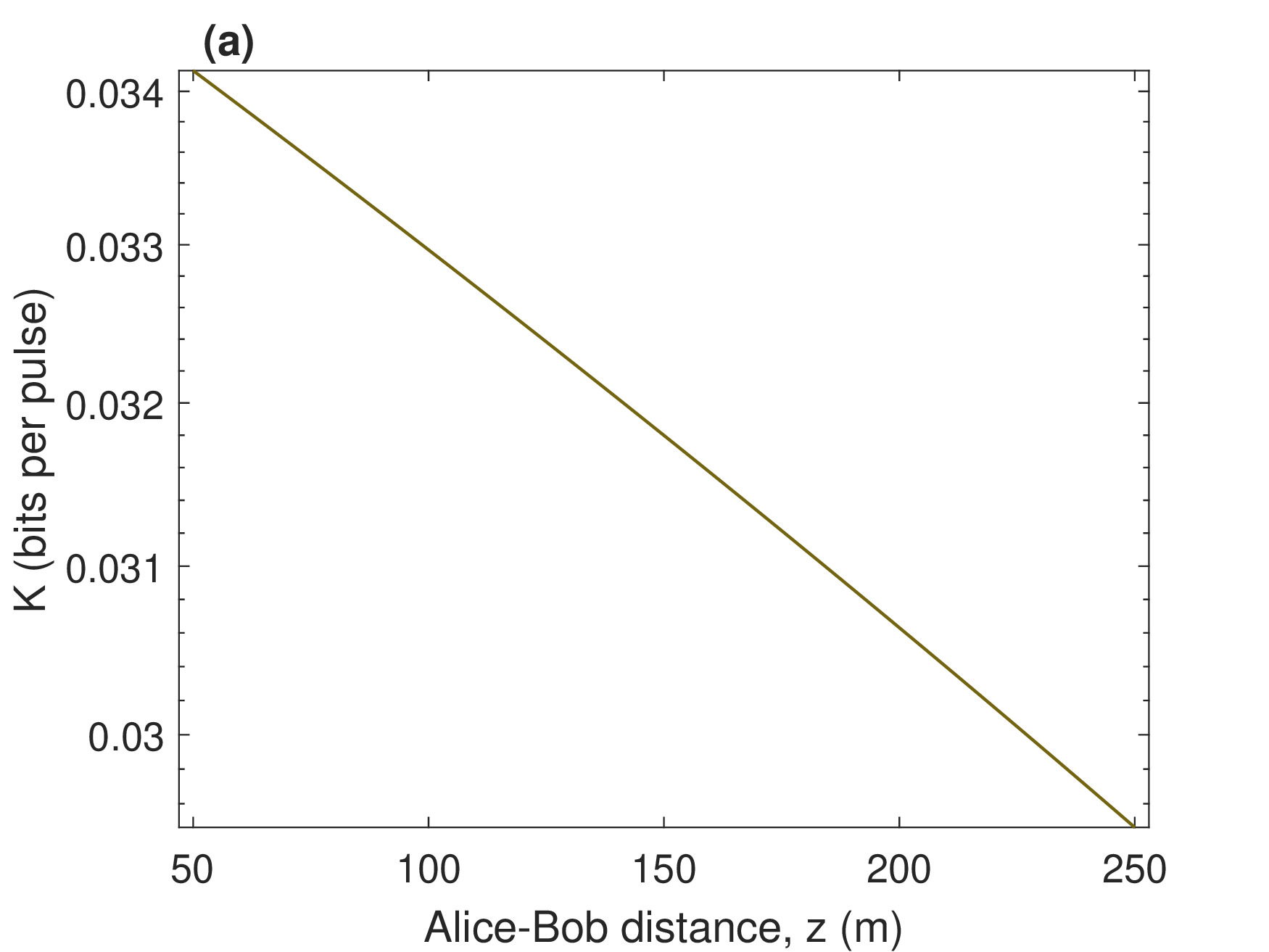}
\includegraphics[width=0.5\textwidth-10pt]{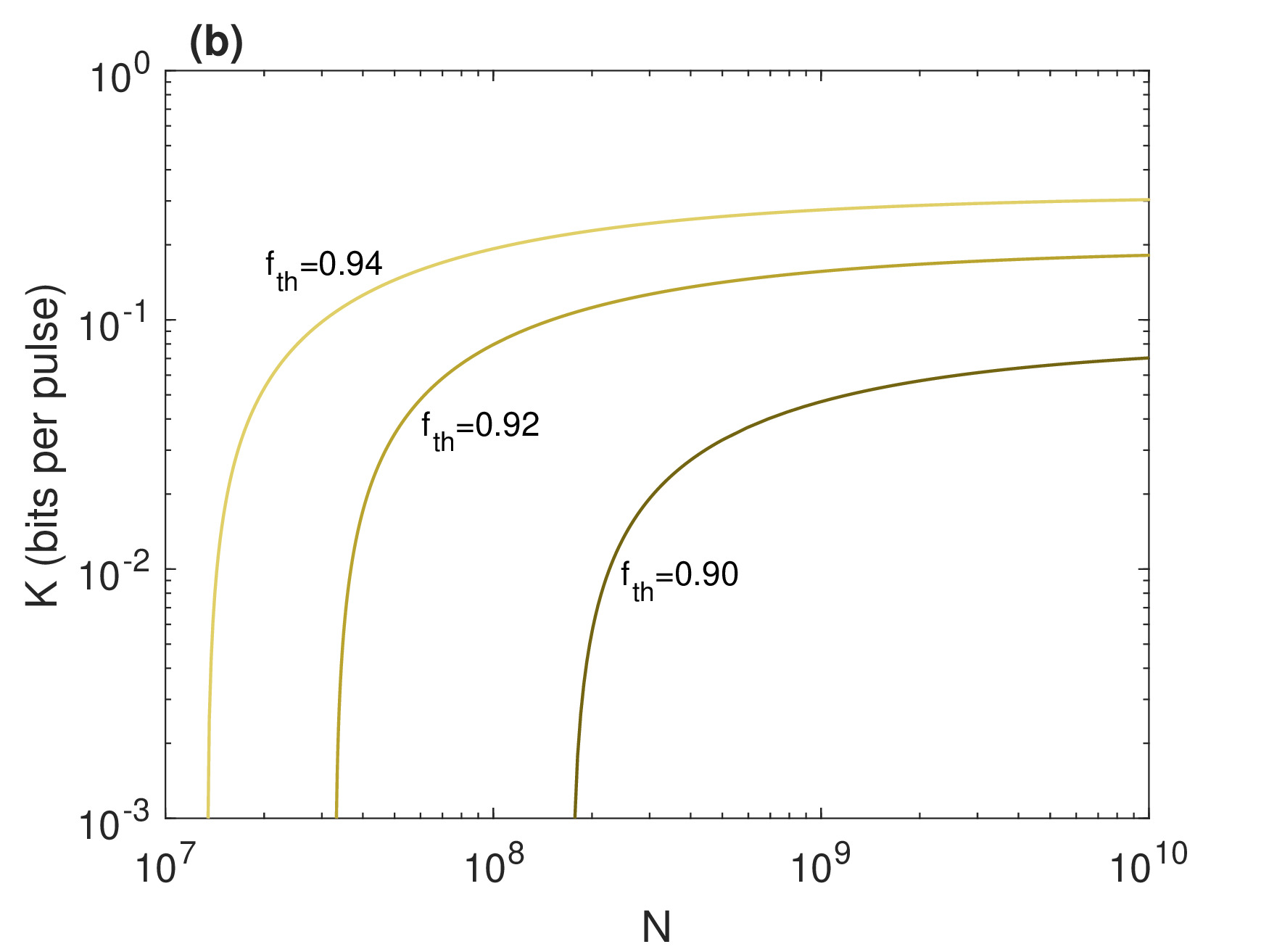}
\caption{{\bf Free-space MDI QKD performance.}
{\bf a,} Secret key rate versus total distance at $N=5\times 10^{8}$ and $f_{\rm th}=0.9$. 
{\bf b,} Secret key rate versus block size at $z=100~\rm m$. 
Set of parameters used are $w_0=10~\rm cm$, $a_{\rm rec}=20~\rm cm$, $\beta=0.98$, $\mu_A=\mu_B=45$, $\eta_A=0.98$, $\eta_{\rm eff}=0.98$, $\nu_{\rm el}=0.01$, $\nu_{\rm det}=1$, $\alpha_0=5\times10^{-6}$, $C_n^2=1.28\times 10^{-14}$, $\overline{n}_{\rm bg}=4.8\times 10^{-12}$, $\delta=10~\rm\mu m$, and $\overline{n}_{\rm ex}=0.04$. 
Other parameters related to pilots and parameter estimation are $m_{\rm PL}=0.1N$, $m_{\rm PE}=0.1N$, $d=2^6$, $\textsc{FER}=0.1$, $\varepsilon_{\rm s}=\varepsilon_{\rm h}=\varepsilon_{\rm pe}=10^{-10}$, $w=6.34$ and $\varepsilon=4.5\times 10^{-10}$. 
Note that realistic block sizes are up to $10^8$ with current data processing facilities.} 
\label{fig:rate}
\end{figure}

Figure~\ref{fig:rate} partly reveals the performance of CV MDI QKD in a free-space setup. 
Here, we assume a horizontal path between Alice and Bob, both located at $H_A=H_B=20~\rm m$. We refer to the caption for the nominal (reasonably realistic) parameters that we have used. 
As we discussed under Fig.~\ref{fig:WeakTB}, we can use Yura's weak turbulence theory. However, let us emphasis that Yura's condition ($\phi\ll 1$) has to be considered approximate as we consider a weak satisfaction of it, i.e., at all distances considered in Fig.~\ref{fig:rate} we have that $\phi<0.4$.

In Fig.~\ref{fig:rate}a we plot the average rate versus distance at fixed block size $N=5\times 10^{8}$ by assuming postselection threshold $f_{\rm th}=0.9$ to build a stable channel. Next, in order to see the effect of block-size, and point out the difference between different values of $f_{\rm th}$, we plot the average rate versus block size at fixed distance $z=100~\rm m$. 
It is observed that smaller values of postselection threshold would result in very poor performance of the system, or it requires a very high, impractical block-size.
For instance, with the same set of parameters given in Fig.~\ref{fig:rate}, the protocol is incapable of delivering a positive rate at $f_{\rm th}=0.85$.

%%%%%%%%%%%%%%%%%%%%%
\section{Slant paths} 
It is conceivable that either of the stations of Alice and Bob is located at a higher altitude than the other, e.g., on top of a mountain. Furthermore, they can be moving objects such as HAPSs. In either case we face a slant atmospheric path between Alice and Bob. 
Supposedly, in such scenarios, the beam light propagates through different atmospheric layers; hence, a more elaborate consideration may be required. For instance, we note that the index-of-refraction structure $C_n^2$ is not anymore constant and changes with the altitude. 

To begin with, let us assume a slant path between a HAPS, say Bob's station at altitude $H_B$, and Alice's platform on the ground, located at $H_A<H_B$ above sea-level. The length of the path is given by 
\begin{align}
    z= & \sqrt{(R_E+H_B)^2+(R_E+H_A)^2(\cos^2\theta-1)} \notag \\ & -(R_E+H_A)\cos\theta,
\end{align}
where $R_E\simeq 6371$~km is earth's radius and $\theta$ the zenith angle. 
As the first consideration, in the following we try to identify the regime of turbulence that is determinant of the choice of equations to be used. 

A more general, altitude-dependent expression for scintillation index, to be used instead of the Rytov variance is~\cite{Andrews:Book,Andrews:2000} 
\begin{align}  
\label{eq:ScintIndex}
   & \sigma_\textsc{si}^2(\theta,H_B)= -1 + \\ 
   & \exp \left[\frac{0.49\beta_\textsc{r}^2(\theta,H_B)}{\Big(1+1.11 \beta_\textsc{r}^{12/5}(\theta,H_B)\Big)^{7/6}} + \frac{0.51\beta_\textsc{r}^2(\theta,H_B)}{\Big(1+0.69 \beta_\textsc{r}^{12/5}(\theta,H_B)\Big)^{5/6}} \right] \notag 
\end{align}
where 
\begin{align}
\beta_\textsc{r}^2(\theta,H_B)=2.25k^{7/6} \sec^{11/6}(\theta) \int_{H_A}^{H_B} dh~ (h-H_A)^{5/6} C_n^2(h),   \notag
\end{align}
and a downlink path is (from Bob to Alice) assumed. 
According to the Hufnagel-Valley (H-V) atmospheric model~\cite[Sec.~12.2]{Andrews:Book}, the index-of-refraction structure is a function of the altitude $h$ and the windspeed $v$
\begin{align}
\label{Cn2h}
    C_n^2(h)= & 5.94\times 10^{-53}(v/27)^2 h^{10} e^{-h/1000} \notag \\ & + 2.7\times 10^{-16}e^{-h/1500}+ Ae^{-h/100},
\end{align}
where $A$ is the nominal value of $C_n^2(0)$ at the ground.  

\begin{figure}[t]
\vspace{+.1cm}
\includegraphics[width=0.5\textwidth-10pt]{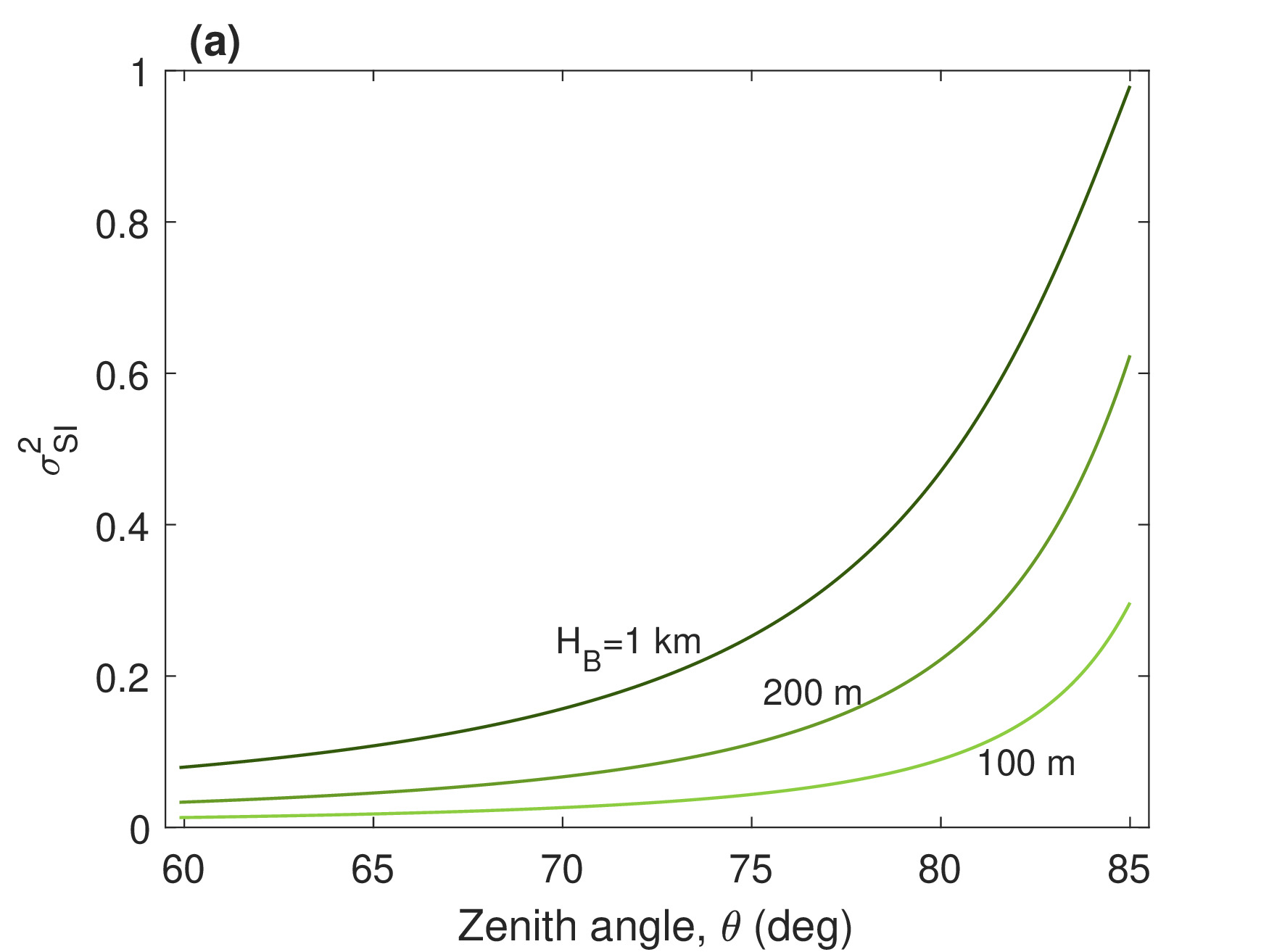}
\includegraphics[width=0.5\textwidth-10pt]{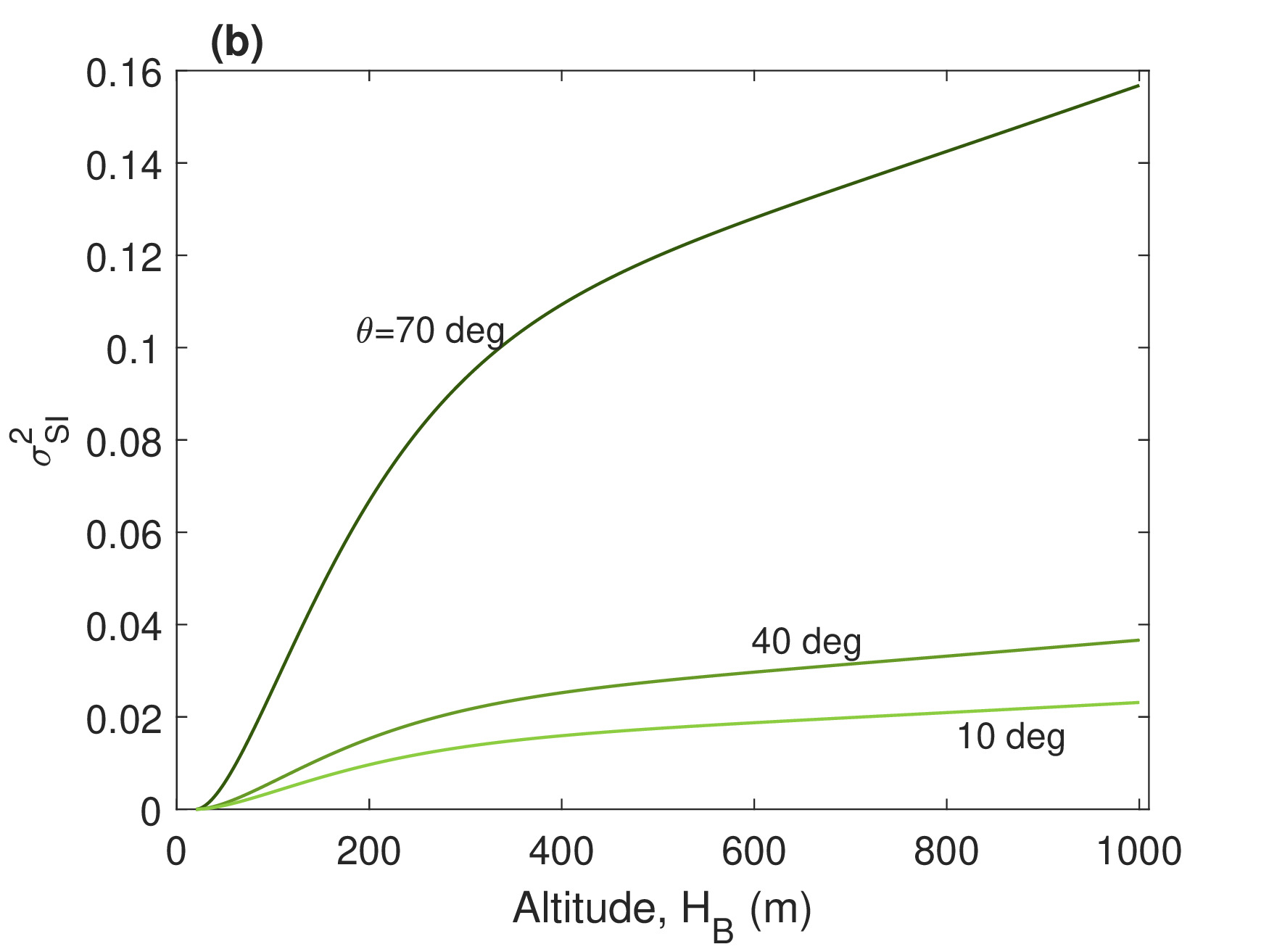}
\caption{{\bf High-altitude platform systems.} 
{\bf a,} Scintillation index versus the zenith angle at different altitudes of Bob's station. 
{\bf b,} Scintillation index versus Bob's altitude at different zenith angles. We assume a fixed of $H_A=20$m for Alice's station. } 
\label{fig:SintInd}
\end{figure}

From Fig.~\ref{fig:SintInd} it is seen that the regime of turbulence can be assumed weak. Here we have considered low-wind, $v=21~\rm ms^{-1}$, and night-time with $A=1.7\times 10^{-14}~\rm m^{-2/3}$~\cite{Pirandola:Sat2021,Andrews:Book}. 
Consequently, in such slant-path regimes, we can still make use of the Yura's recipe for a weak turbulent atmosphere. Let us now get back to our MDI QKD protocol and apply the above considerations to the analysis. 

\begin{figure}[b]
\vspace{+.1cm}
\includegraphics[width=0.5\textwidth-10pt]{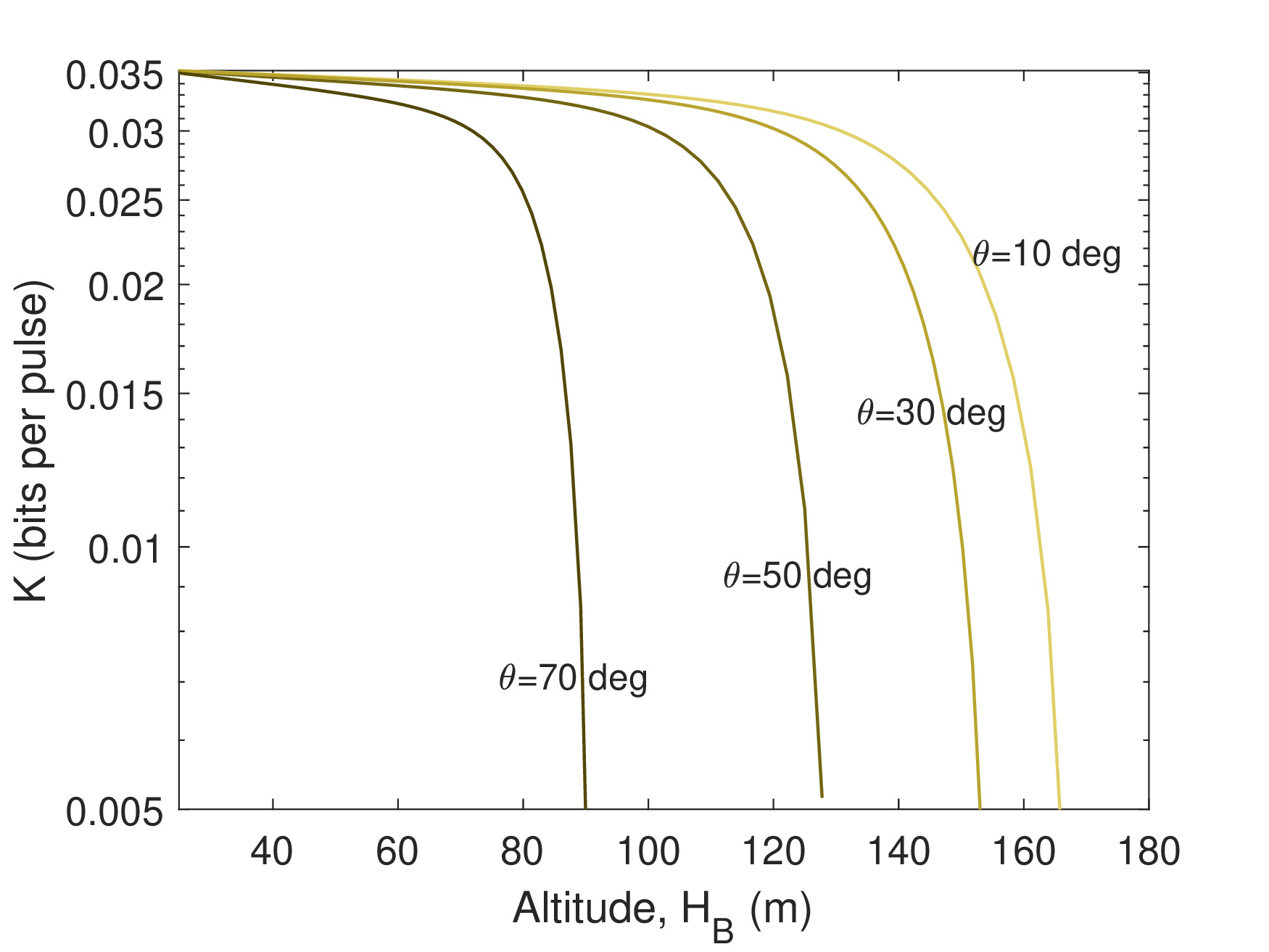}
\caption{{\bf Free-space MDI QKD performance with high-altitude platform systems.} 
Secret key rate versus Bob's altitude at different zenith angles with $H_A=20~\rm m$ at all cases. The curves represent the rate at $f_{\rm th}=0.9$ and the set of parameters used here are the same as reported in Fig.~\ref{fig:rate}, except for  the index-of-refraction structure $C_n^2$ which is varying.} 
\label{fig:rate_SP}
\end{figure}

The performance of CV MDI QKD with slant paths can be seen in Fig.~\ref{fig:rate_SP}, where for several values of zenith angle we have plotted composable finite-size key rate at night-time operation.   
Here we have set the same parameters as given in Fig.~\ref{fig:rate}, including initial beam size $w_0=10$~cm, receiver size $a_{\rm rec}=20$~cm, block size $N=5\times 10^8$, and pilot postselection threshold $f_{\rm th}=0.9$.   
Our simulation illustrates that with a reasonable block size and receiver size quantum communications through CV MDI protocols is feasible for altitudes up to $H_B=200$~m (note that Alice's altitude is fixed at $H_A=20$~m).

%%%%%%%%%%%%%%%%%
\section{Summary}

In the present work, we have developed a composable security analysis of CV-MDI-QKD over free-space optical links that can include several types of noise and experimental inefficiencies. 
We have demonstrated that asymmetric CV MDI QKD protocols can be used to extract a composably-secure key over FSO links. 
This can be achieved in the powerful collective eavesdropping scenario with the protocol offering substantially high rates. 
We have considered physical space-related phenomena such as light-beam diffraction, deflection, turbulence, and beam widening, all of which degrade transmissivity. We have also accounted for several types of noise, including background noise, excess noise and receiver noise, that free-space CV QKD suffers from. 
Furthermore, we have studied the usefulness of the protocol for slant path through an atmospheric turbulent space. 
In all cases we show that high-rate CV-MDI-QKD is possible over short FSO links of the order of hundred meters, where the regime of turbulence is weak.

\section*{Acknowledgments}
This work has been funded by the EPSRC via the UK Quantum Communications Hub (Grant No. EP/T001011/1).
SP would like to thank G. Mortzou.

%%%%%%%%%%%%%%%%%%%%%%%%%
\bibliography{references}

\begin{thebibliography}{49}%
\makeatletter
\providecommand \@ifxundefined [1]{%
 \@ifx{#1\undefined}
}%
\providecommand \@ifnum [1]{%
 \ifnum #1\expandafter \@firstoftwo
 \else \expandafter \@secondoftwo
 \fi
}%
\providecommand \@ifx [1]{%
 \ifx #1\expandafter \@firstoftwo
 \else \expandafter \@secondoftwo
 \fi
}%
\providecommand \natexlab [1]{#1}%
\providecommand \enquote  [1]{``#1''}%
\providecommand \bibnamefont  [1]{#1}%
\providecommand \bibfnamefont [1]{#1}%
\providecommand \citenamefont [1]{#1}%
\providecommand \href@noop [0]{\@secondoftwo}%
\providecommand \href [0]{\begingroup \@sanitize@url \@href}%
\providecommand \@href[1]{\@@startlink{#1}\@@href}%
\providecommand \@@href[1]{\endgroup#1\@@endlink}%
\providecommand \@sanitize@url [0]{\catcode `\\12\catcode `\$12\catcode
  `\&12\catcode `\#12\catcode `\^12\catcode `\_12\catcode `\%12\relax}%
\providecommand \@@startlink[1]{}%
\providecommand \@@endlink[0]{}%
\providecommand \url  [0]{\begingroup\@sanitize@url \@url }%
\providecommand \@url [1]{\endgroup\@href {#1}{\urlprefix }}%
\providecommand \urlprefix  [0]{URL }%
\providecommand \Eprint [0]{\href }%
\providecommand \doibase [0]{http://dx.doi.org/}%
\providecommand \selectlanguage [0]{\@gobble}%
\providecommand \bibinfo  [0]{\@secondoftwo}%
\providecommand \bibfield  [0]{\@secondoftwo}%
\providecommand \translation [1]{[#1]}%
\providecommand \BibitemOpen [0]{}%
\providecommand \bibitemStop [0]{}%
\providecommand \bibitemNoStop [0]{.\EOS\space}%
\providecommand \EOS [0]{\spacefactor3000\relax}%
\providecommand \BibitemShut  [1]{\csname bibitem#1\endcsname}%
\let\auto@bib@innerbib\@empty
%</preamble>
\bibitem [{\citenamefont {Pirandola}\ \emph {et~al.}(2020)\citenamefont
  {Pirandola}, \citenamefont {Andersen}, \citenamefont {Banchi}, \citenamefont
  {Berta}, \citenamefont {Bunandar}, \citenamefont {Colbeck}, \citenamefont
  {Englund}, \citenamefont {Gehring}, \citenamefont {Lupo}, \citenamefont
  {Ottaviani}, \citenamefont {Pereira}, \citenamefont {Razavi}, \citenamefont
  {Shaari}, \citenamefont {Tomamichel}, \citenamefont {Usenko}, \citenamefont
  {Vallone}, \citenamefont {Villoresi},\ and\ \citenamefont
  {Wallden}}]{Pirandola:AQCrypt}%
  \BibitemOpen
  \bibfield  {author} {\bibinfo {author} {\bibfnamefont {S.}~\bibnamefont
  {Pirandola}}, \bibinfo {author} {\bibfnamefont {U.~L.}\ \bibnamefont
  {Andersen}}, \bibinfo {author} {\bibfnamefont {L.}~\bibnamefont {Banchi}},
  \bibinfo {author} {\bibfnamefont {M.}~\bibnamefont {Berta}}, \bibinfo
  {author} {\bibfnamefont {D.}~\bibnamefont {Bunandar}}, \bibinfo {author}
  {\bibfnamefont {R.}~\bibnamefont {Colbeck}}, \bibinfo {author} {\bibfnamefont
  {D.}~\bibnamefont {Englund}}, \bibinfo {author} {\bibfnamefont
  {T.}~\bibnamefont {Gehring}}, \bibinfo {author} {\bibfnamefont
  {C.}~\bibnamefont {Lupo}}, \bibinfo {author} {\bibfnamefont {C.}~\bibnamefont
  {Ottaviani}}, \bibinfo {author} {\bibfnamefont {J.~L.}\ \bibnamefont
  {Pereira}}, \bibinfo {author} {\bibfnamefont {M.}~\bibnamefont {Razavi}},
  \bibinfo {author} {\bibfnamefont {J.~S.}\ \bibnamefont {Shaari}}, \bibinfo
  {author} {\bibfnamefont {M.}~\bibnamefont {Tomamichel}}, \bibinfo {author}
  {\bibfnamefont {V.~C.}\ \bibnamefont {Usenko}}, \bibinfo {author}
  {\bibfnamefont {G.}~\bibnamefont {Vallone}}, \bibinfo {author} {\bibfnamefont
  {P.}~\bibnamefont {Villoresi}}, \ and\ \bibinfo {author} {\bibfnamefont
  {P.}~\bibnamefont {Wallden}},\ }\href {\doibase 10.1364/AOP.361502}
  {\bibfield  {journal} {\bibinfo  {journal} {Adv. Opt. Photon.}\ }\textbf
  {\bibinfo {volume} {12}},\ \bibinfo {pages} {1012} (\bibinfo {year}
  {2020})}\BibitemShut {NoStop}%
\bibitem [{\citenamefont {Gisin}\ \emph {et~al.}(2002)\citenamefont {Gisin},
  \citenamefont {Ribordy}, \citenamefont {Tittel},\ and\ \citenamefont
  {Zbinden}}]{Gisin:Rev2002}%
  \BibitemOpen
  \bibfield  {author} {\bibinfo {author} {\bibfnamefont {N.}~\bibnamefont
  {Gisin}}, \bibinfo {author} {\bibfnamefont {G.}~\bibnamefont {Ribordy}},
  \bibinfo {author} {\bibfnamefont {W.}~\bibnamefont {Tittel}}, \ and\ \bibinfo
  {author} {\bibfnamefont {H.}~\bibnamefont {Zbinden}},\ }\href {\doibase
  10.1103/RevModPhys.74.145} {\bibfield  {journal} {\bibinfo  {journal} {Rev.
  Mod. Phys.}\ }\textbf {\bibinfo {volume} {74}},\ \bibinfo {pages} {145}
  (\bibinfo {year} {2002})}\BibitemShut {NoStop}%
\bibitem [{\citenamefont {Pirandola}\ \emph {et~al.}(2017)\citenamefont
  {Pirandola}, \citenamefont {Laurenza}, \citenamefont {Ottaviani},\ and\
  \citenamefont {Banchi}}]{Pirandola:PLOB2017}%
  \BibitemOpen
  \bibfield  {author} {\bibinfo {author} {\bibfnamefont {S.}~\bibnamefont
  {Pirandola}}, \bibinfo {author} {\bibfnamefont {R.}~\bibnamefont {Laurenza}},
  \bibinfo {author} {\bibfnamefont {C.}~\bibnamefont {Ottaviani}}, \ and\
  \bibinfo {author} {\bibfnamefont {L.}~\bibnamefont {Banchi}},\ }\href
  {\doibase https://doi.org/10.1038/ncomms15043} {\bibfield  {journal}
  {\bibinfo  {journal} {Nat. Commun.}\ }\textbf {\bibinfo {volume} {8}},\
  \bibinfo {pages} {15043} (\bibinfo {year} {2017})}\BibitemShut {NoStop}%
\bibitem [{\citenamefont {Briegel}\ \emph {et~al.}(1998)\citenamefont
  {Briegel}, \citenamefont {D\"ur}, \citenamefont {Cirac},\ and\ \citenamefont
  {Zoller}}]{Briegel:QRs1998}%
  \BibitemOpen
  \bibfield  {author} {\bibinfo {author} {\bibfnamefont {H.-J.}\ \bibnamefont
  {Briegel}}, \bibinfo {author} {\bibfnamefont {W.}~\bibnamefont {D\"ur}},
  \bibinfo {author} {\bibfnamefont {J.~I.}\ \bibnamefont {Cirac}}, \ and\
  \bibinfo {author} {\bibfnamefont {P.}~\bibnamefont {Zoller}},\ }\href
  {\doibase 10.1103/PhysRevLett.81.5932} {\bibfield  {journal} {\bibinfo
  {journal} {Phys. Rev. Lett.}\ }\textbf {\bibinfo {volume} {81}},\ \bibinfo
  {pages} {5932} (\bibinfo {year} {1998})}\BibitemShut {NoStop}%
\bibitem [{\citenamefont {D\"ur}\ \emph {et~al.}(1999)\citenamefont {D\"ur},
  \citenamefont {Briegel}, \citenamefont {Cirac},\ and\ \citenamefont
  {Zoller}}]{Dur:QRs1999}%
  \BibitemOpen
  \bibfield  {author} {\bibinfo {author} {\bibfnamefont {W.}~\bibnamefont
  {D\"ur}}, \bibinfo {author} {\bibfnamefont {H.-J.}\ \bibnamefont {Briegel}},
  \bibinfo {author} {\bibfnamefont {J.~I.}\ \bibnamefont {Cirac}}, \ and\
  \bibinfo {author} {\bibfnamefont {P.}~\bibnamefont {Zoller}},\ }\href
  {\doibase 10.1103/PhysRevA.59.169} {\bibfield  {journal} {\bibinfo  {journal}
  {Phys. Rev. A}\ }\textbf {\bibinfo {volume} {59}},\ \bibinfo {pages} {169}
  (\bibinfo {year} {1999})}\BibitemShut {NoStop}%
\bibitem [{\citenamefont {Dür}\ \emph {et~al.}(2016)\citenamefont {Dür},
  \citenamefont {Briegel}, \citenamefont {Zoller},\ and\ \citenamefont
  {Loock}}]{Dur:QRs2016}%
  \BibitemOpen
  \bibfield  {author} {\bibinfo {author} {\bibfnamefont {W.}~\bibnamefont
  {Dür}}, \bibinfo {author} {\bibfnamefont {H.-J.}\ \bibnamefont {Briegel}},
  \bibinfo {author} {\bibfnamefont {P.}~\bibnamefont {Zoller}}, \ and\ \bibinfo
  {author} {\bibfnamefont {P.~v.}\ \bibnamefont {Loock}},\ }\enquote {\bibinfo
  {title} {Quantum repeater},}\ in\ \href {\doibase
  https://doi.org/10.1002/9783527805785.ch30} {\emph {\bibinfo {booktitle}
  {Quantum Information}}}\ (\bibinfo  {publisher} {John Wiley \& Sons, Ltd},\
  \bibinfo {year} {2016})\ Chap.~\bibinfo {chapter} {30}, pp.\ \bibinfo {pages}
  {691--700}\BibitemShut {NoStop}%
\bibitem [{\citenamefont {Furrer}\ and\ \citenamefont
  {Munro}(2018)}]{Furrer:QRs2018}%
  \BibitemOpen
  \bibfield  {author} {\bibinfo {author} {\bibfnamefont {F.}~\bibnamefont
  {Furrer}}\ and\ \bibinfo {author} {\bibfnamefont {W.~J.}\ \bibnamefont
  {Munro}},\ }\href {\doibase 10.1103/PhysRevA.98.032335} {\bibfield  {journal}
  {\bibinfo  {journal} {Phys. Rev. A}\ }\textbf {\bibinfo {volume} {98}},\
  \bibinfo {pages} {032335} (\bibinfo {year} {2018})}\BibitemShut {NoStop}%
\bibitem [{\citenamefont {Dias}\ \emph {et~al.}(2020)\citenamefont {Dias},
  \citenamefont {Winnel}, \citenamefont {Hosseinidehaj},\ and\ \citenamefont
  {Ralph}}]{Dias:QRs2020}%
  \BibitemOpen
  \bibfield  {author} {\bibinfo {author} {\bibfnamefont {J.}~\bibnamefont
  {Dias}}, \bibinfo {author} {\bibfnamefont {M.~S.}\ \bibnamefont {Winnel}},
  \bibinfo {author} {\bibfnamefont {N.}~\bibnamefont {Hosseinidehaj}}, \ and\
  \bibinfo {author} {\bibfnamefont {T.~C.}\ \bibnamefont {Ralph}},\ }\href
  {\doibase 10.1103/PhysRevA.102.052425} {\bibfield  {journal} {\bibinfo
  {journal} {Phys. Rev. A}\ }\textbf {\bibinfo {volume} {102}},\ \bibinfo
  {pages} {052425} (\bibinfo {year} {2020})}\BibitemShut {NoStop}%
\bibitem [{\citenamefont {Seshadreesan}\ \emph {et~al.}(2020)\citenamefont
  {Seshadreesan}, \citenamefont {Krovi},\ and\ \citenamefont
  {Guha}}]{Seshadreesan:QRs2020}%
  \BibitemOpen
  \bibfield  {author} {\bibinfo {author} {\bibfnamefont {K.~P.}\ \bibnamefont
  {Seshadreesan}}, \bibinfo {author} {\bibfnamefont {H.}~\bibnamefont {Krovi}},
  \ and\ \bibinfo {author} {\bibfnamefont {S.}~\bibnamefont {Guha}},\ }\href
  {\doibase 10.1103/PhysRevResearch.2.013310} {\bibfield  {journal} {\bibinfo
  {journal} {Phys. Rev. Research}\ }\textbf {\bibinfo {volume} {2}},\ \bibinfo
  {pages} {013310} (\bibinfo {year} {2020})}\BibitemShut {NoStop}%
\bibitem [{\citenamefont {Ghalaii}\ and\ \citenamefont
  {Pirandola}(2020)}]{Ghalaii:QRs2020}%
  \BibitemOpen
  \bibfield  {author} {\bibinfo {author} {\bibfnamefont {M.}~\bibnamefont
  {Ghalaii}}\ and\ \bibinfo {author} {\bibfnamefont {S.}~\bibnamefont
  {Pirandola}},\ }\href {\doibase 10.1103/PhysRevA.102.062412} {\bibfield
  {journal} {\bibinfo  {journal} {Phys. Rev. A}\ }\textbf {\bibinfo {volume}
  {102}},\ \bibinfo {pages} {062412} (\bibinfo {year} {2020})}\BibitemShut
  {NoStop}%
\bibitem [{\citenamefont {Liao}\ \emph {et~al.}(2017)\citenamefont {Liao},
  \citenamefont {Cai}, \citenamefont {Liu}, \citenamefont {Zhang},
  \citenamefont {Li}, \citenamefont {Ren}, \citenamefont {Yin}, \citenamefont
  {Shen}, \citenamefont {Cao}, \citenamefont {Li}, \citenamefont {Li},
  \citenamefont {Chen}, \citenamefont {Sun}, \citenamefont {Jia}, \citenamefont
  {Wu}, \citenamefont {Jiang}, \citenamefont {Wang}, \citenamefont {Huang},
  \citenamefont {Wang}, \citenamefont {Zhou}, \citenamefont {Deng},
  \citenamefont {Xi}, \citenamefont {Ma}, \citenamefont {Hu}, \citenamefont
  {Zhang}, \citenamefont {Chen}, \citenamefont {Liu}, \citenamefont {Wang},
  \citenamefont {Zhu}, \citenamefont {Lu}, \citenamefont {Shu}, \citenamefont
  {Peng}, \citenamefont {Wang},\ and\ \citenamefont {Pan}}]{Liao:Nat2017}%
  \BibitemOpen
  \bibfield  {author} {\bibinfo {author} {\bibfnamefont {S.-K.}\ \bibnamefont
  {Liao}}, \bibinfo {author} {\bibfnamefont {W.-Q.}\ \bibnamefont {Cai}},
  \bibinfo {author} {\bibfnamefont {W.-Y.}\ \bibnamefont {Liu}}, \bibinfo
  {author} {\bibfnamefont {L.}~\bibnamefont {Zhang}}, \bibinfo {author}
  {\bibfnamefont {Y.}~\bibnamefont {Li}}, \bibinfo {author} {\bibfnamefont
  {J.-G.}\ \bibnamefont {Ren}}, \bibinfo {author} {\bibfnamefont
  {J.}~\bibnamefont {Yin}}, \bibinfo {author} {\bibfnamefont {Q.}~\bibnamefont
  {Shen}}, \bibinfo {author} {\bibfnamefont {Y.}~\bibnamefont {Cao}}, \bibinfo
  {author} {\bibfnamefont {Z.-P.}\ \bibnamefont {Li}}, \bibinfo {author}
  {\bibfnamefont {F.-Z.}\ \bibnamefont {Li}}, \bibinfo {author} {\bibfnamefont
  {X.-W.}\ \bibnamefont {Chen}}, \bibinfo {author} {\bibfnamefont {L.-H.}\
  \bibnamefont {Sun}}, \bibinfo {author} {\bibfnamefont {J.-J.}\ \bibnamefont
  {Jia}}, \bibinfo {author} {\bibfnamefont {J.-C.}\ \bibnamefont {Wu}},
  \bibinfo {author} {\bibfnamefont {X.-J.}\ \bibnamefont {Jiang}}, \bibinfo
  {author} {\bibfnamefont {J.-F.}\ \bibnamefont {Wang}}, \bibinfo {author}
  {\bibfnamefont {Y.-M.}\ \bibnamefont {Huang}}, \bibinfo {author}
  {\bibfnamefont {Q.}~\bibnamefont {Wang}}, \bibinfo {author} {\bibfnamefont
  {Y.-L.}\ \bibnamefont {Zhou}}, \bibinfo {author} {\bibfnamefont
  {L.}~\bibnamefont {Deng}}, \bibinfo {author} {\bibfnamefont {T.}~\bibnamefont
  {Xi}}, \bibinfo {author} {\bibfnamefont {L.}~\bibnamefont {Ma}}, \bibinfo
  {author} {\bibfnamefont {T.}~\bibnamefont {Hu}}, \bibinfo {author}
  {\bibfnamefont {Q.}~\bibnamefont {Zhang}}, \bibinfo {author} {\bibfnamefont
  {Y.-A.}\ \bibnamefont {Chen}}, \bibinfo {author} {\bibfnamefont {N.-L.}\
  \bibnamefont {Liu}}, \bibinfo {author} {\bibfnamefont {X.-B.}\ \bibnamefont
  {Wang}}, \bibinfo {author} {\bibfnamefont {Z.-C.}\ \bibnamefont {Zhu}},
  \bibinfo {author} {\bibfnamefont {C.-Y.}\ \bibnamefont {Lu}}, \bibinfo
  {author} {\bibfnamefont {R.}~\bibnamefont {Shu}}, \bibinfo {author}
  {\bibfnamefont {C.-Z.}\ \bibnamefont {Peng}}, \bibinfo {author}
  {\bibfnamefont {J.-Y.}\ \bibnamefont {Wang}}, \ and\ \bibinfo {author}
  {\bibfnamefont {J.-W.}\ \bibnamefont {Pan}},\ }\href {\doibase
  10.1038/nature23655} {\bibfield  {journal} {\bibinfo  {journal} {Nature}\
  }\textbf {\bibinfo {volume} {549}},\ \bibinfo {pages} {43} (\bibinfo {year}
  {2017})}\BibitemShut {NoStop}%
\bibitem [{\citenamefont {Ren}\ \emph {et~al.}(2017)\citenamefont {Ren},
  \citenamefont {Xu}, \citenamefont {Yong}, \citenamefont {Zhang},
  \citenamefont {Liao}, \citenamefont {Yin}, \citenamefont {Liu}, \citenamefont
  {Cai}, \citenamefont {Yang}, \citenamefont {Li}, \citenamefont {Yang},
  \citenamefont {Han}, \citenamefont {Yao}, \citenamefont {Li}, \citenamefont
  {Wu}, \citenamefont {Wan}, \citenamefont {Liu}, \citenamefont {Liu},
  \citenamefont {Kuang}, \citenamefont {He}, \citenamefont {Shang},
  \citenamefont {Guo}, \citenamefont {Zheng}, \citenamefont {Tian},
  \citenamefont {Zhu}, \citenamefont {Liu}, \citenamefont {Lu}, \citenamefont
  {Shu}, \citenamefont {Chen}, \citenamefont {Peng}, \citenamefont {Wang},\
  and\ \citenamefont {Pan}}]{Ren:Nat2017}%
  \BibitemOpen
  \bibfield  {author} {\bibinfo {author} {\bibfnamefont {J.-G.}\ \bibnamefont
  {Ren}}, \bibinfo {author} {\bibfnamefont {P.}~\bibnamefont {Xu}}, \bibinfo
  {author} {\bibfnamefont {H.-L.}\ \bibnamefont {Yong}}, \bibinfo {author}
  {\bibfnamefont {L.}~\bibnamefont {Zhang}}, \bibinfo {author} {\bibfnamefont
  {S.-K.}\ \bibnamefont {Liao}}, \bibinfo {author} {\bibfnamefont
  {J.}~\bibnamefont {Yin}}, \bibinfo {author} {\bibfnamefont {W.-Y.}\
  \bibnamefont {Liu}}, \bibinfo {author} {\bibfnamefont {W.-Q.}\ \bibnamefont
  {Cai}}, \bibinfo {author} {\bibfnamefont {M.}~\bibnamefont {Yang}}, \bibinfo
  {author} {\bibfnamefont {L.}~\bibnamefont {Li}}, \bibinfo {author}
  {\bibfnamefont {K.-X.}\ \bibnamefont {Yang}}, \bibinfo {author}
  {\bibfnamefont {X.}~\bibnamefont {Han}}, \bibinfo {author} {\bibfnamefont
  {Y.-Q.}\ \bibnamefont {Yao}}, \bibinfo {author} {\bibfnamefont
  {J.}~\bibnamefont {Li}}, \bibinfo {author} {\bibfnamefont {H.-Y.}\
  \bibnamefont {Wu}}, \bibinfo {author} {\bibfnamefont {S.}~\bibnamefont
  {Wan}}, \bibinfo {author} {\bibfnamefont {L.}~\bibnamefont {Liu}}, \bibinfo
  {author} {\bibfnamefont {D.-Q.}\ \bibnamefont {Liu}}, \bibinfo {author}
  {\bibfnamefont {Y.-W.}\ \bibnamefont {Kuang}}, \bibinfo {author}
  {\bibfnamefont {Z.-P.}\ \bibnamefont {He}}, \bibinfo {author} {\bibfnamefont
  {P.}~\bibnamefont {Shang}}, \bibinfo {author} {\bibfnamefont
  {C.}~\bibnamefont {Guo}}, \bibinfo {author} {\bibfnamefont {R.-H.}\
  \bibnamefont {Zheng}}, \bibinfo {author} {\bibfnamefont {K.}~\bibnamefont
  {Tian}}, \bibinfo {author} {\bibfnamefont {Z.-C.}\ \bibnamefont {Zhu}},
  \bibinfo {author} {\bibfnamefont {N.-L.}\ \bibnamefont {Liu}}, \bibinfo
  {author} {\bibfnamefont {C.-Y.}\ \bibnamefont {Lu}}, \bibinfo {author}
  {\bibfnamefont {R.}~\bibnamefont {Shu}}, \bibinfo {author} {\bibfnamefont
  {Y.-A.}\ \bibnamefont {Chen}}, \bibinfo {author} {\bibfnamefont {C.-Z.}\
  \bibnamefont {Peng}}, \bibinfo {author} {\bibfnamefont {J.-Y.}\ \bibnamefont
  {Wang}}, \ and\ \bibinfo {author} {\bibfnamefont {J.-W.}\ \bibnamefont
  {Pan}},\ }\href {\doibase 10.1038/nature23675} {\bibfield  {journal}
  {\bibinfo  {journal} {Nature}\ }\textbf {\bibinfo {volume} {549}},\ \bibinfo
  {pages} {70} (\bibinfo {year} {2017})}\BibitemShut {NoStop}%
\bibitem [{\citenamefont {Yin}\ \emph {et~al.}(2017)\citenamefont {Yin},
  \citenamefont {Cao}, \citenamefont {Li}, \citenamefont {Liao}, \citenamefont
  {Zhang}, \citenamefont {Ren}, \citenamefont {Cai}, \citenamefont {Liu},
  \citenamefont {Li}, \citenamefont {Dai}, \citenamefont {Li}, \citenamefont
  {Lu}, \citenamefont {Gong}, \citenamefont {Xu}, \citenamefont {Li},
  \citenamefont {Li}, \citenamefont {Yin}, \citenamefont {Jiang}, \citenamefont
  {Li}, \citenamefont {Jia}, \citenamefont {Ren}, \citenamefont {He},
  \citenamefont {Zhou}, \citenamefont {Zhang}, \citenamefont {Wang},
  \citenamefont {Chang}, \citenamefont {Zhu}, \citenamefont {Liu},
  \citenamefont {Chen}, \citenamefont {Lu}, \citenamefont {Shu}, \citenamefont
  {Peng}, \citenamefont {Wang},\ and\ \citenamefont {Pan}}]{Yin:1200kmSat}%
  \BibitemOpen
  \bibfield  {author} {\bibinfo {author} {\bibfnamefont {J.}~\bibnamefont
  {Yin}}, \bibinfo {author} {\bibfnamefont {Y.}~\bibnamefont {Cao}}, \bibinfo
  {author} {\bibfnamefont {Y.-H.}\ \bibnamefont {Li}}, \bibinfo {author}
  {\bibfnamefont {S.-K.}\ \bibnamefont {Liao}}, \bibinfo {author}
  {\bibfnamefont {L.}~\bibnamefont {Zhang}}, \bibinfo {author} {\bibfnamefont
  {J.-G.}\ \bibnamefont {Ren}}, \bibinfo {author} {\bibfnamefont {W.-Q.}\
  \bibnamefont {Cai}}, \bibinfo {author} {\bibfnamefont {W.-Y.}\ \bibnamefont
  {Liu}}, \bibinfo {author} {\bibfnamefont {B.}~\bibnamefont {Li}}, \bibinfo
  {author} {\bibfnamefont {H.}~\bibnamefont {Dai}}, \bibinfo {author}
  {\bibfnamefont {G.-B.}\ \bibnamefont {Li}}, \bibinfo {author} {\bibfnamefont
  {Q.-M.}\ \bibnamefont {Lu}}, \bibinfo {author} {\bibfnamefont {Y.-H.}\
  \bibnamefont {Gong}}, \bibinfo {author} {\bibfnamefont {Y.}~\bibnamefont
  {Xu}}, \bibinfo {author} {\bibfnamefont {S.-L.}\ \bibnamefont {Li}}, \bibinfo
  {author} {\bibfnamefont {F.-Z.}\ \bibnamefont {Li}}, \bibinfo {author}
  {\bibfnamefont {Y.-Y.}\ \bibnamefont {Yin}}, \bibinfo {author} {\bibfnamefont
  {Z.-Q.}\ \bibnamefont {Jiang}}, \bibinfo {author} {\bibfnamefont
  {M.}~\bibnamefont {Li}}, \bibinfo {author} {\bibfnamefont {J.-J.}\
  \bibnamefont {Jia}}, \bibinfo {author} {\bibfnamefont {G.}~\bibnamefont
  {Ren}}, \bibinfo {author} {\bibfnamefont {D.}~\bibnamefont {He}}, \bibinfo
  {author} {\bibfnamefont {Y.-L.}\ \bibnamefont {Zhou}}, \bibinfo {author}
  {\bibfnamefont {X.-X.}\ \bibnamefont {Zhang}}, \bibinfo {author}
  {\bibfnamefont {N.}~\bibnamefont {Wang}}, \bibinfo {author} {\bibfnamefont
  {X.}~\bibnamefont {Chang}}, \bibinfo {author} {\bibfnamefont {Z.-C.}\
  \bibnamefont {Zhu}}, \bibinfo {author} {\bibfnamefont {N.-L.}\ \bibnamefont
  {Liu}}, \bibinfo {author} {\bibfnamefont {Y.-A.}\ \bibnamefont {Chen}},
  \bibinfo {author} {\bibfnamefont {C.-Y.}\ \bibnamefont {Lu}}, \bibinfo
  {author} {\bibfnamefont {R.}~\bibnamefont {Shu}}, \bibinfo {author}
  {\bibfnamefont {C.-Z.}\ \bibnamefont {Peng}}, \bibinfo {author}
  {\bibfnamefont {J.-Y.}\ \bibnamefont {Wang}}, \ and\ \bibinfo {author}
  {\bibfnamefont {J.-W.}\ \bibnamefont {Pan}},\ }\href {\doibase
  10.1126/science.aan3211} {\bibfield  {journal} {\bibinfo  {journal}
  {Science}\ }\textbf {\bibinfo {volume} {356}},\ \bibinfo {pages} {1140}
  (\bibinfo {year} {2017})}\BibitemShut {NoStop}%
\bibitem [{\citenamefont {Bedington}\ \emph {et~al.}(2016)\citenamefont
  {Bedington}, \citenamefont {Bai}, \citenamefont {Truong-Cao}, \citenamefont
  {Tan}, \citenamefont {Durak}, \citenamefont {Zafra}, \citenamefont {Grieve},
  \citenamefont {Oi},\ and\ \citenamefont {Ling}}]{Bedington:NSat2016}%
  \BibitemOpen
  \bibfield  {author} {\bibinfo {author} {\bibfnamefont {R.}~\bibnamefont
  {Bedington}}, \bibinfo {author} {\bibfnamefont {X.}~\bibnamefont {Bai}},
  \bibinfo {author} {\bibfnamefont {E.}~\bibnamefont {Truong-Cao}}, \bibinfo
  {author} {\bibfnamefont {Y.~C.}\ \bibnamefont {Tan}}, \bibinfo {author}
  {\bibfnamefont {K.}~\bibnamefont {Durak}}, \bibinfo {author} {\bibfnamefont
  {A.~V.}\ \bibnamefont {Zafra}}, \bibinfo {author} {\bibfnamefont {J.~A.}\
  \bibnamefont {Grieve}}, \bibinfo {author} {\bibfnamefont {D.~K.}\
  \bibnamefont {Oi}}, \ and\ \bibinfo {author} {\bibfnamefont {A.}~\bibnamefont
  {Ling}},\ }\href {https://doi.org/10.1140/epjqt/s40507-016-0051-7} {\bibfield
   {journal} {\bibinfo  {journal} {EPJ Quantum Technology}\ }\textbf {\bibinfo
  {volume} {3}},\ \bibinfo {pages} {12} (\bibinfo {year} {2016})}\BibitemShut
  {NoStop}%
\bibitem [{\citenamefont {Dequal}\ \emph {et~al.}(2021)\citenamefont {Dequal},
  \citenamefont {Vidarte}, \citenamefont {Rodriguez}, \citenamefont {Vallone},
  \citenamefont {Villoresi}, \citenamefont {Leverrier},\ and\ \citenamefont
  {Diamanti}}]{Dequal:njp2021}%
  \BibitemOpen
  \bibfield  {author} {\bibinfo {author} {\bibfnamefont {D.}~\bibnamefont
  {Dequal}}, \bibinfo {author} {\bibfnamefont {L.~T.}\ \bibnamefont {Vidarte}},
  \bibinfo {author} {\bibfnamefont {V.~R.}\ \bibnamefont {Rodriguez}}, \bibinfo
  {author} {\bibfnamefont {G.}~\bibnamefont {Vallone}}, \bibinfo {author}
  {\bibfnamefont {P.}~\bibnamefont {Villoresi}}, \bibinfo {author}
  {\bibfnamefont {A.}~\bibnamefont {Leverrier}}, \ and\ \bibinfo {author}
  {\bibfnamefont {E.}~\bibnamefont {Diamanti}},\ }\href {\doibase
  10.1038/s41534-020-00336-4} {\bibfield  {journal} {\bibinfo  {journal} {npj
  Quantum Inf}\ }\textbf {\bibinfo {volume} {7}},\ \bibinfo {pages} {3}
  (\bibinfo {year} {2021})}\BibitemShut {NoStop}%
\bibitem [{\citenamefont {Mazzarella}\ \emph {et~al.}(2020)\citenamefont
  {Mazzarella}, \citenamefont {Lowe}, \citenamefont {Lowndes}, \citenamefont
  {Joshi}, \citenamefont {Greenland}, \citenamefont {McNeil}, \citenamefont
  {Mercury}, \citenamefont {Macdonald}, \citenamefont {Rarity},\ and\
  \citenamefont {Oi}}]{Mazzarella:QUARC2020}%
  \BibitemOpen
  \bibfield  {author} {\bibinfo {author} {\bibfnamefont {L.}~\bibnamefont
  {Mazzarella}}, \bibinfo {author} {\bibfnamefont {C.}~\bibnamefont {Lowe}},
  \bibinfo {author} {\bibfnamefont {D.}~\bibnamefont {Lowndes}}, \bibinfo
  {author} {\bibfnamefont {S.~K.}\ \bibnamefont {Joshi}}, \bibinfo {author}
  {\bibfnamefont {S.}~\bibnamefont {Greenland}}, \bibinfo {author}
  {\bibfnamefont {D.}~\bibnamefont {McNeil}}, \bibinfo {author} {\bibfnamefont
  {C.}~\bibnamefont {Mercury}}, \bibinfo {author} {\bibfnamefont
  {M.}~\bibnamefont {Macdonald}}, \bibinfo {author} {\bibfnamefont
  {J.}~\bibnamefont {Rarity}}, \ and\ \bibinfo {author} {\bibfnamefont
  {D.~K.~L.}\ \bibnamefont {Oi}},\ }\href {\doibase
  10.3390/cryptography4010007} {\bibfield  {journal} {\bibinfo  {journal}
  {Cryptography}\ }\textbf {\bibinfo {volume} {4(1)}},\ \bibinfo {pages} {7}
  (\bibinfo {year} {2020})}\BibitemShut {NoStop}%
\bibitem [{\citenamefont {Sidhu}\ \emph {et~al.}(2021)\citenamefont {Sidhu},
  \citenamefont {Joshi}, \citenamefont {Gündoğan}, \citenamefont {Brougham},
  \citenamefont {Lowndes}, \citenamefont {Mazzarella}, \citenamefont {Krutzik},
  \citenamefont {Mohapatra}, \citenamefont {Dequal}, \citenamefont {Vallone},
  \citenamefont {Villoresi}, \citenamefont {Ling}, \citenamefont {Jennewein},
  \citenamefont {Mohageg}, \citenamefont {Rarity}, \citenamefont {Fuentes},
  \citenamefont {Pirandola},\ and\ \citenamefont {Oi}}]{Sidhu:ASpQC}%
  \BibitemOpen
  \bibfield  {author} {\bibinfo {author} {\bibfnamefont {J.~S.}\ \bibnamefont
  {Sidhu}}, \bibinfo {author} {\bibfnamefont {S.~K.}\ \bibnamefont {Joshi}},
  \bibinfo {author} {\bibfnamefont {M.}~\bibnamefont {Gündoğan}}, \bibinfo
  {author} {\bibfnamefont {T.}~\bibnamefont {Brougham}}, \bibinfo {author}
  {\bibfnamefont {D.}~\bibnamefont {Lowndes}}, \bibinfo {author} {\bibfnamefont
  {L.}~\bibnamefont {Mazzarella}}, \bibinfo {author} {\bibfnamefont
  {M.}~\bibnamefont {Krutzik}}, \bibinfo {author} {\bibfnamefont
  {S.}~\bibnamefont {Mohapatra}}, \bibinfo {author} {\bibfnamefont
  {D.}~\bibnamefont {Dequal}}, \bibinfo {author} {\bibfnamefont
  {G.}~\bibnamefont {Vallone}}, \bibinfo {author} {\bibfnamefont
  {P.}~\bibnamefont {Villoresi}}, \bibinfo {author} {\bibfnamefont
  {A.}~\bibnamefont {Ling}}, \bibinfo {author} {\bibfnamefont {T.}~\bibnamefont
  {Jennewein}}, \bibinfo {author} {\bibfnamefont {M.}~\bibnamefont {Mohageg}},
  \bibinfo {author} {\bibfnamefont {J.~G.}\ \bibnamefont {Rarity}}, \bibinfo
  {author} {\bibfnamefont {I.}~\bibnamefont {Fuentes}}, \bibinfo {author}
  {\bibfnamefont {S.}~\bibnamefont {Pirandola}}, \ and\ \bibinfo {author}
  {\bibfnamefont {D.~K.~L.}\ \bibnamefont {Oi}},\ }\href@noop {} {\bibfield
  {journal} {\bibinfo  {journal} {IET Quantum Communication}\ } (\bibinfo
  {year} {2021})}\BibitemShut {NoStop}%
\bibitem [{\citenamefont {Zhang}\ \emph {et~al.}(2018)\citenamefont {Zhang},
  \citenamefont {Xu}, \citenamefont {Chen}, \citenamefont {Peng},\ and\
  \citenamefont {Pan}}]{Zhang:Optica2018}%
  \BibitemOpen
  \bibfield  {author} {\bibinfo {author} {\bibfnamefont {Q.}~\bibnamefont
  {Zhang}}, \bibinfo {author} {\bibfnamefont {F.}~\bibnamefont {Xu}}, \bibinfo
  {author} {\bibfnamefont {Y.-A.}\ \bibnamefont {Chen}}, \bibinfo {author}
  {\bibfnamefont {C.-Z.}\ \bibnamefont {Peng}}, \ and\ \bibinfo {author}
  {\bibfnamefont {J.-W.}\ \bibnamefont {Pan}},\ }\href {\doibase
  10.1364/OE.26.024260} {\bibfield  {journal} {\bibinfo  {journal} {Opt.
  Express}\ }\textbf {\bibinfo {volume} {26}},\ \bibinfo {pages} {24260}
  (\bibinfo {year} {2018})}\BibitemShut {NoStop}%
\bibitem [{\citenamefont {Kimble}(2008)}]{Kimble:QuInternet}%
  \BibitemOpen
  \bibfield  {author} {\bibinfo {author} {\bibfnamefont {H.~J.}\ \bibnamefont
  {Kimble}},\ }\href {\doibase 10.1038/nature07127} {\bibfield  {journal}
  {\bibinfo  {journal} {Nature}\ }\textbf {\bibinfo {volume} {453}},\ \bibinfo
  {pages} {1023} (\bibinfo {year} {2008})}\BibitemShut {NoStop}%
\bibitem [{\citenamefont {Pirandola}\ and\ \citenamefont
  {Braunstein}(2016)}]{Pirandola:QInternet}%
  \BibitemOpen
  \bibfield  {author} {\bibinfo {author} {\bibfnamefont {S.}~\bibnamefont
  {Pirandola}}\ and\ \bibinfo {author} {\bibfnamefont {S.~L.}\ \bibnamefont
  {Braunstein}},\ }\href {\doibase 10.1038/532169a} {\bibfield  {journal}
  {\bibinfo  {journal} {Nature}\ }\textbf {\bibinfo {volume} {532}},\ \bibinfo
  {pages} {169} (\bibinfo {year} {2016})}\BibitemShut {NoStop}%
\bibitem [{\citenamefont {Pirandola}(2021{\natexlab{a}})}]{Pirandola:FS2021}%
  \BibitemOpen
  \bibfield  {author} {\bibinfo {author} {\bibfnamefont {S.}~\bibnamefont
  {Pirandola}},\ }\href {\doibase 10.1103/PhysRevResearch.3.013279} {\bibfield
  {journal} {\bibinfo  {journal} {Phys. Rev. Research}\ }\textbf {\bibinfo
  {volume} {3}},\ \bibinfo {pages} {013279} (\bibinfo {year}
  {2021}{\natexlab{a}})}\BibitemShut {NoStop}%
\bibitem [{\citenamefont {Pirandola}(2021{\natexlab{b}})}]{Pirandola:Sat2021}%
  \BibitemOpen
  \bibfield  {author} {\bibinfo {author} {\bibfnamefont {S.}~\bibnamefont
  {Pirandola}},\ }\href {\doibase 10.1103/PhysRevResearch.3.023130} {\bibfield
  {journal} {\bibinfo  {journal} {Phys. Rev. Research}\ }\textbf {\bibinfo
  {volume} {3}},\ \bibinfo {pages} {023130} (\bibinfo {year}
  {2021}{\natexlab{b}})}\BibitemShut {NoStop}%
\bibitem [{\citenamefont
  {Pirandola}(2021{\natexlab{c}})}]{Pirandola:CompCVQKD2021}%
  \BibitemOpen
  \bibfield  {author} {\bibinfo {author} {\bibfnamefont {S.}~\bibnamefont
  {Pirandola}},\ }\href {\doibase 10.1103/PhysRevResearch.3.043014} {\bibfield
  {journal} {\bibinfo  {journal} {Phys. Rev. Research}\ }\textbf {\bibinfo
  {volume} {3}},\ \bibinfo {pages} {043014} (\bibinfo {year}
  {2021}{\natexlab{c}})}\BibitemShut {NoStop}%
\bibitem [{\citenamefont {Ghalaii}\ and\ \citenamefont
  {Pirandola}(2022)}]{Ghalaii:StTurb2021}%
  \BibitemOpen
  \bibfield  {author} {\bibinfo {author} {\bibfnamefont {M.}~\bibnamefont
  {Ghalaii}}\ and\ \bibinfo {author} {\bibfnamefont {S.}~\bibnamefont
  {Pirandola}},\ }\href {\doibase 10.1038/s42005-022-00814-5} {\bibfield
  {journal} {\bibinfo  {journal} {Communications Physics}\ }\textbf {\bibinfo
  {volume} {5}},\ \bibinfo {pages} {38} (\bibinfo {year} {2022})}\BibitemShut
  {NoStop}%
\bibitem [{\citenamefont {Braunstein}\ and\ \citenamefont
  {Pirandola}(2012)}]{Braunstein:SideCh2012}%
  \BibitemOpen
  \bibfield  {author} {\bibinfo {author} {\bibfnamefont {S.~L.}\ \bibnamefont
  {Braunstein}}\ and\ \bibinfo {author} {\bibfnamefont {S.}~\bibnamefont
  {Pirandola}},\ }\href {\doibase 10.1103/PhysRevLett.108.130502} {\bibfield
  {journal} {\bibinfo  {journal} {Phys. Rev. Lett.}\ }\textbf {\bibinfo
  {volume} {108}},\ \bibinfo {pages} {130502} (\bibinfo {year}
  {2012})}\BibitemShut {NoStop}%
\bibitem [{\citenamefont {Lo}\ \emph {et~al.}(2012)\citenamefont {Lo},
  \citenamefont {Curty},\ and\ \citenamefont {Qi}}]{Lo:MDI2012}%
  \BibitemOpen
  \bibfield  {author} {\bibinfo {author} {\bibfnamefont {H.-K.}\ \bibnamefont
  {Lo}}, \bibinfo {author} {\bibfnamefont {M.}~\bibnamefont {Curty}}, \ and\
  \bibinfo {author} {\bibfnamefont {B.}~\bibnamefont {Qi}},\ }\href {\doibase
  10.1103/PhysRevLett.108.130503} {\bibfield  {journal} {\bibinfo  {journal}
  {Phys. Rev. Lett.}\ }\textbf {\bibinfo {volume} {108}},\ \bibinfo {pages}
  {130503} (\bibinfo {year} {2012})}\BibitemShut {NoStop}%
\bibitem [{\citenamefont {Rubenok}\ \emph {et~al.}(2013)\citenamefont
  {Rubenok}, \citenamefont {Slater}, \citenamefont {Chan}, \citenamefont
  {Lucio-Martinez},\ and\ \citenamefont {Tittel}}]{Rubenok:DVMDI2013}%
  \BibitemOpen
  \bibfield  {author} {\bibinfo {author} {\bibfnamefont {A.}~\bibnamefont
  {Rubenok}}, \bibinfo {author} {\bibfnamefont {J.~A.}\ \bibnamefont {Slater}},
  \bibinfo {author} {\bibfnamefont {P.}~\bibnamefont {Chan}}, \bibinfo {author}
  {\bibfnamefont {I.}~\bibnamefont {Lucio-Martinez}}, \ and\ \bibinfo {author}
  {\bibfnamefont {W.}~\bibnamefont {Tittel}},\ }\href {\doibase
  10.1103/PhysRevLett.111.130501} {\bibfield  {journal} {\bibinfo  {journal}
  {Phys. Rev. Lett.}\ }\textbf {\bibinfo {volume} {111}},\ \bibinfo {pages}
  {130501} (\bibinfo {year} {2013})}\BibitemShut {NoStop}%
\bibitem [{\citenamefont {Liu}\ \emph {et~al.}(2013)\citenamefont {Liu},
  \citenamefont {Chen}, \citenamefont {Wang}, \citenamefont {Liang},
  \citenamefont {Shentu}, \citenamefont {Wang}, \citenamefont {Cui},
  \citenamefont {Yin}, \citenamefont {Liu}, \citenamefont {Li}, \citenamefont
  {Ma}, \citenamefont {Pelc}, \citenamefont {Fejer}, \citenamefont {Peng},
  \citenamefont {Zhang},\ and\ \citenamefont {Pan}}]{Liu:DVMDI2013}%
  \BibitemOpen
  \bibfield  {author} {\bibinfo {author} {\bibfnamefont {Y.}~\bibnamefont
  {Liu}}, \bibinfo {author} {\bibfnamefont {T.-Y.}\ \bibnamefont {Chen}},
  \bibinfo {author} {\bibfnamefont {L.-J.}\ \bibnamefont {Wang}}, \bibinfo
  {author} {\bibfnamefont {H.}~\bibnamefont {Liang}}, \bibinfo {author}
  {\bibfnamefont {G.-L.}\ \bibnamefont {Shentu}}, \bibinfo {author}
  {\bibfnamefont {J.}~\bibnamefont {Wang}}, \bibinfo {author} {\bibfnamefont
  {K.}~\bibnamefont {Cui}}, \bibinfo {author} {\bibfnamefont {H.-L.}\
  \bibnamefont {Yin}}, \bibinfo {author} {\bibfnamefont {N.-L.}\ \bibnamefont
  {Liu}}, \bibinfo {author} {\bibfnamefont {L.}~\bibnamefont {Li}}, \bibinfo
  {author} {\bibfnamefont {X.}~\bibnamefont {Ma}}, \bibinfo {author}
  {\bibfnamefont {J.~S.}\ \bibnamefont {Pelc}}, \bibinfo {author}
  {\bibfnamefont {M.~M.}\ \bibnamefont {Fejer}}, \bibinfo {author}
  {\bibfnamefont {C.-Z.}\ \bibnamefont {Peng}}, \bibinfo {author}
  {\bibfnamefont {Q.}~\bibnamefont {Zhang}}, \ and\ \bibinfo {author}
  {\bibfnamefont {J.-W.}\ \bibnamefont {Pan}},\ }\href {\doibase
  10.1103/PhysRevLett.111.130502} {\bibfield  {journal} {\bibinfo  {journal}
  {Phys. Rev. Lett.}\ }\textbf {\bibinfo {volume} {111}},\ \bibinfo {pages}
  {130502} (\bibinfo {year} {2013})}\BibitemShut {NoStop}%
\bibitem [{\citenamefont {Ferreira~da Silva}\ \emph {et~al.}(2013)\citenamefont
  {Ferreira~da Silva}, \citenamefont {Vitoreti}, \citenamefont {Xavier},
  \citenamefont {do~Amaral}, \citenamefont {Tempor\~ao},\ and\ \citenamefont
  {von~der Weid}}]{daSilva:DVMDI2013}%
  \BibitemOpen
  \bibfield  {author} {\bibinfo {author} {\bibfnamefont {T.}~\bibnamefont
  {Ferreira~da Silva}}, \bibinfo {author} {\bibfnamefont {D.}~\bibnamefont
  {Vitoreti}}, \bibinfo {author} {\bibfnamefont {G.~B.}\ \bibnamefont
  {Xavier}}, \bibinfo {author} {\bibfnamefont {G.~C.}\ \bibnamefont
  {do~Amaral}}, \bibinfo {author} {\bibfnamefont {G.~P.}\ \bibnamefont
  {Tempor\~ao}}, \ and\ \bibinfo {author} {\bibfnamefont {J.~P.}\ \bibnamefont
  {von~der Weid}},\ }\href {\doibase 10.1103/PhysRevA.88.052303} {\bibfield
  {journal} {\bibinfo  {journal} {Phys. Rev. A}\ }\textbf {\bibinfo {volume}
  {88}},\ \bibinfo {pages} {052303} (\bibinfo {year} {2013})}\BibitemShut
  {NoStop}%
\bibitem [{\citenamefont {Cao}\ \emph {et~al.}(2020)\citenamefont {Cao},
  \citenamefont {Li}, \citenamefont {Yang}, \citenamefont {Jiang},
  \citenamefont {Li}, \citenamefont {Hu}, \citenamefont {Abulizi},
  \citenamefont {Li}, \citenamefont {Zhang}, \citenamefont {Sun}, \citenamefont
  {Liu}, \citenamefont {Jiang}, \citenamefont {Liao}, \citenamefont {Ren},
  \citenamefont {Li}, \citenamefont {You}, \citenamefont {Wang}, \citenamefont
  {Yin}, \citenamefont {Lu}, \citenamefont {Wang}, \citenamefont {Zhang},
  \citenamefont {Peng},\ and\ \citenamefont {Pan}}]{Cao:FSDVMDI2020}%
  \BibitemOpen
  \bibfield  {author} {\bibinfo {author} {\bibfnamefont {Y.}~\bibnamefont
  {Cao}}, \bibinfo {author} {\bibfnamefont {Y.-H.}\ \bibnamefont {Li}},
  \bibinfo {author} {\bibfnamefont {K.-X.}\ \bibnamefont {Yang}}, \bibinfo
  {author} {\bibfnamefont {Y.-F.}\ \bibnamefont {Jiang}}, \bibinfo {author}
  {\bibfnamefont {S.-L.}\ \bibnamefont {Li}}, \bibinfo {author} {\bibfnamefont
  {X.-L.}\ \bibnamefont {Hu}}, \bibinfo {author} {\bibfnamefont
  {M.}~\bibnamefont {Abulizi}}, \bibinfo {author} {\bibfnamefont {C.-L.}\
  \bibnamefont {Li}}, \bibinfo {author} {\bibfnamefont {W.}~\bibnamefont
  {Zhang}}, \bibinfo {author} {\bibfnamefont {Q.-C.}\ \bibnamefont {Sun}},
  \bibinfo {author} {\bibfnamefont {W.-Y.}\ \bibnamefont {Liu}}, \bibinfo
  {author} {\bibfnamefont {X.}~\bibnamefont {Jiang}}, \bibinfo {author}
  {\bibfnamefont {S.-K.}\ \bibnamefont {Liao}}, \bibinfo {author}
  {\bibfnamefont {J.-G.}\ \bibnamefont {Ren}}, \bibinfo {author} {\bibfnamefont
  {H.}~\bibnamefont {Li}}, \bibinfo {author} {\bibfnamefont {L.}~\bibnamefont
  {You}}, \bibinfo {author} {\bibfnamefont {Z.}~\bibnamefont {Wang}}, \bibinfo
  {author} {\bibfnamefont {J.}~\bibnamefont {Yin}}, \bibinfo {author}
  {\bibfnamefont {C.-Y.}\ \bibnamefont {Lu}}, \bibinfo {author} {\bibfnamefont
  {X.-B.}\ \bibnamefont {Wang}}, \bibinfo {author} {\bibfnamefont
  {Q.}~\bibnamefont {Zhang}}, \bibinfo {author} {\bibfnamefont {C.-Z.}\
  \bibnamefont {Peng}}, \ and\ \bibinfo {author} {\bibfnamefont {J.-W.}\
  \bibnamefont {Pan}},\ }\href {\doibase 10.1103/PhysRevLett.125.260503}
  {\bibfield  {journal} {\bibinfo  {journal} {Phys. Rev. Lett.}\ }\textbf
  {\bibinfo {volume} {125}},\ \bibinfo {pages} {260503} (\bibinfo {year}
  {2020})}\BibitemShut {NoStop}%
\bibitem [{\citenamefont {Wang}\ \emph
  {et~al.}(2021{\natexlab{a}})\citenamefont {Wang}, \citenamefont {Dong},
  \citenamefont {Zhao}, \citenamefont {Liu}, \citenamefont {Liu},\ and\
  \citenamefont {Zhu}}]{Wang:FSDVMDI2021a}%
  \BibitemOpen
  \bibfield  {author} {\bibinfo {author} {\bibfnamefont {X.}~\bibnamefont
  {Wang}}, \bibinfo {author} {\bibfnamefont {C.}~\bibnamefont {Dong}}, \bibinfo
  {author} {\bibfnamefont {S.}~\bibnamefont {Zhao}}, \bibinfo {author}
  {\bibfnamefont {Y.}~\bibnamefont {Liu}}, \bibinfo {author} {\bibfnamefont
  {X.}~\bibnamefont {Liu}}, \ and\ \bibinfo {author} {\bibfnamefont
  {H.}~\bibnamefont {Zhu}},\ }\href {\doibase 10.1088/1367-2630/abf534}
  {\bibfield  {journal} {\bibinfo  {journal} {New Journal of Physics}\ }\textbf
  {\bibinfo {volume} {23}},\ \bibinfo {pages} {045001} (\bibinfo {year}
  {2021}{\natexlab{a}})}\BibitemShut {NoStop}%
\bibitem [{\citenamefont {Wang}\ \emph
  {et~al.}(2021{\natexlab{b}})\citenamefont {Wang}, \citenamefont {Liu},
  \citenamefont {Wu}, \citenamefont {Guo}, \citenamefont {Zhang}, \citenamefont
  {Zhao},\ and\ \citenamefont {Dong}}]{Wang:FSDVMDI2021b}%
  \BibitemOpen
  \bibfield  {author} {\bibinfo {author} {\bibfnamefont {X.}~\bibnamefont
  {Wang}}, \bibinfo {author} {\bibfnamefont {W.}~\bibnamefont {Liu}}, \bibinfo
  {author} {\bibfnamefont {T.}~\bibnamefont {Wu}}, \bibinfo {author}
  {\bibfnamefont {C.}~\bibnamefont {Guo}}, \bibinfo {author} {\bibfnamefont
  {Y.}~\bibnamefont {Zhang}}, \bibinfo {author} {\bibfnamefont
  {S.}~\bibnamefont {Zhao}}, \ and\ \bibinfo {author} {\bibfnamefont
  {C.}~\bibnamefont {Dong}},\ }\href {\doibase 10.3390/e23101299} {\bibfield
  {journal} {\bibinfo  {journal} {Entropy}\ }\textbf {\bibinfo {volume} {23}}
  (\bibinfo {year} {2021}{\natexlab{b}}),\ 10.3390/e23101299}\BibitemShut
  {NoStop}%
\bibitem [{\citenamefont {Dong}\ \emph {et~al.}(2021)\citenamefont {Dong},
  \citenamefont {Huang}, \citenamefont {Cui},\ and\ \citenamefont
  {Jiao}}]{Dong:FSDVMDI2021}%
  \BibitemOpen
  \bibfield  {author} {\bibinfo {author} {\bibfnamefont {Q.}~\bibnamefont
  {Dong}}, \bibinfo {author} {\bibfnamefont {G.}~\bibnamefont {Huang}},
  \bibinfo {author} {\bibfnamefont {W.}~\bibnamefont {Cui}}, \ and\ \bibinfo
  {author} {\bibfnamefont {R.}~\bibnamefont {Jiao}},\ }\href {\doibase
  10.1088/2058-9565/ac37b2} {\bibfield  {journal} {\bibinfo  {journal} {Quantum
  Science and Technology}\ }\textbf {\bibinfo {volume} {7}},\ \bibinfo {pages}
  {015014} (\bibinfo {year} {2021})}\BibitemShut {NoStop}%
\bibitem [{\citenamefont {Pirandola}\ \emph {et~al.}(2013)\citenamefont
  {Pirandola}, \citenamefont {Ottaviani}, \citenamefont {Spedalieri},
  \citenamefont {Weedbrook},\ and\ \citenamefont
  {Braunstein}}]{Pirandola:CV2013}%
  \BibitemOpen
  \bibfield  {author} {\bibinfo {author} {\bibfnamefont {S.}~\bibnamefont
  {Pirandola}}, \bibinfo {author} {\bibfnamefont {C.}~\bibnamefont
  {Ottaviani}}, \bibinfo {author} {\bibfnamefont {G.}~\bibnamefont
  {Spedalieri}}, \bibinfo {author} {\bibfnamefont {C.}~\bibnamefont
  {Weedbrook}}, \ and\ \bibinfo {author} {\bibfnamefont {S.~L.}\ \bibnamefont
  {Braunstein}},\ }\href@noop {} {\  (\bibinfo {year} {2013})},\ \Eprint
  {http://arxiv.org/abs/1312.4104v1} {arXiv:1312.4104v1} \BibitemShut {NoStop}%
\bibitem [{\citenamefont {Pirandola}\ \emph {et~al.}(2015)\citenamefont
  {Pirandola}, \citenamefont {Ottaviani}, \citenamefont {Spedalieri},
  \citenamefont {Weedbrook}, \citenamefont {Braunstein}, \citenamefont {Lloyd},
  \citenamefont {Gehring}, \citenamefont {Jacobsen},\ and\ \citenamefont
  {Andersen}}]{Pirandola:CVMDI2015}%
  \BibitemOpen
  \bibfield  {author} {\bibinfo {author} {\bibfnamefont {S.}~\bibnamefont
  {Pirandola}}, \bibinfo {author} {\bibfnamefont {C.}~\bibnamefont
  {Ottaviani}}, \bibinfo {author} {\bibfnamefont {G.}~\bibnamefont
  {Spedalieri}}, \bibinfo {author} {\bibfnamefont {C.}~\bibnamefont
  {Weedbrook}}, \bibinfo {author} {\bibfnamefont {S.~L.}\ \bibnamefont
  {Braunstein}}, \bibinfo {author} {\bibfnamefont {S.}~\bibnamefont {Lloyd}},
  \bibinfo {author} {\bibfnamefont {T.}~\bibnamefont {Gehring}}, \bibinfo
  {author} {\bibfnamefont {C.~S.}\ \bibnamefont {Jacobsen}}, \ and\ \bibinfo
  {author} {\bibfnamefont {U.~L.}\ \bibnamefont {Andersen}},\ }\href {\doibase
  10.1038/nphoton.2015.83} {\bibfield  {journal} {\bibinfo  {journal} {Nature
  Photonics}\ }\textbf {\bibinfo {volume} {9}},\ \bibinfo {pages} {397}
  (\bibinfo {year} {2015})}\BibitemShut {NoStop}%
\bibitem [{\citenamefont {Duntley}(1948)}]{Duntley:1948}%
  \BibitemOpen
  \bibfield  {author} {\bibinfo {author} {\bibfnamefont {S.~Q.}\ \bibnamefont
  {Duntley}},\ }\href {\doibase 10.1364/JOSA.38.000179} {\bibfield  {journal}
  {\bibinfo  {journal} {J. Opt. Soc. Am.}\ }\textbf {\bibinfo {volume} {38}},\
  \bibinfo {pages} {179} (\bibinfo {year} {1948})}\BibitemShut {NoStop}%
\bibitem [{\citenamefont {Bohren}\ and\ \citenamefont
  {Huffman}(2008)}]{Bohren:Book}%
  \BibitemOpen
  \bibfield  {author} {\bibinfo {author} {\bibfnamefont {C.~F.}\ \bibnamefont
  {Bohren}}\ and\ \bibinfo {author} {\bibfnamefont {D.~R.}\ \bibnamefont
  {Huffman}},\ }\href@noop {} {\emph {\bibinfo {title} {Absorption and
  scattering of light by small particles}}}\ (\bibinfo  {publisher} {John Wiley
  \& Sons Inc.},\ \bibinfo {year} {2008})\BibitemShut {NoStop}%
\bibitem [{\citenamefont {Rytov}(1937)}]{Rytov:1937}%
  \BibitemOpen
  \bibfield  {author} {\bibinfo {author} {\bibfnamefont {S.~M.}\ \bibnamefont
  {Rytov}},\ }\href@noop {} {\bibfield  {journal} {\bibinfo  {journal}
  {Izvestiya Akademii Nauk SSSR, Seriya Fizicheskaya (Bulletin of the Academy
  of Sciences of the USSR, Physical Series)}\ }\textbf {\bibinfo {volume}
  {2}},\ \bibinfo {pages} {223} (\bibinfo {year} {1937})}\BibitemShut {NoStop}%
\bibitem [{\citenamefont {Andrews}\ and\ \citenamefont
  {Phillips}(2005)}]{Andrews:Book}%
  \BibitemOpen
  \bibfield  {author} {\bibinfo {author} {\bibfnamefont {L.~C.}\ \bibnamefont
  {Andrews}}\ and\ \bibinfo {author} {\bibfnamefont {R.~L.}\ \bibnamefont
  {Phillips}},\ }\href@noop {} {\emph {\bibinfo {title} {Laser Beam Propagation
  Through Random Medium}}},\ \bibinfo {edition} {2nd}\ ed.\ (\bibinfo
  {publisher} {SPIE},\ \bibinfo {year} {2005})\BibitemShut {NoStop}%
\bibitem [{\citenamefont {Dowling}\ and\ \citenamefont
  {Livingston}(1973)}]{Dowling:BeamWandering}%
  \BibitemOpen
  \bibfield  {author} {\bibinfo {author} {\bibfnamefont {J.~A.}\ \bibnamefont
  {Dowling}}\ and\ \bibinfo {author} {\bibfnamefont {P.~M.}\ \bibnamefont
  {Livingston}},\ }\href {\doibase 10.1364/JOSA.63.000846} {\bibfield
  {journal} {\bibinfo  {journal} {J. Opt. Soc. Am.}\ }\textbf {\bibinfo
  {volume} {63}},\ \bibinfo {pages} {846} (\bibinfo {year} {1973})}\BibitemShut
  {NoStop}%
\bibitem [{\citenamefont {Avesani}\ \emph {et~al.}(2021)\citenamefont
  {Avesani}, \citenamefont {Calderaro}, \citenamefont {Schiavon}, \citenamefont
  {Stanco}, \citenamefont {Agnesi}, \citenamefont {Santamato}, \citenamefont
  {Zahidy}, \citenamefont {Scriminich}, \citenamefont {Foletto}, \citenamefont
  {Contestabile}, \citenamefont {Chiesa}, \citenamefont {Rotta}, \citenamefont
  {Artiglia}, \citenamefont {Montanaro}, \citenamefont {Romagnoli},
  \citenamefont {Sorianello}, \citenamefont {Vedovato}, \citenamefont
  {Vallone},\ and\ \citenamefont {Villoresi}}]{Avesani:2021}%
  \BibitemOpen
  \bibfield  {author} {\bibinfo {author} {\bibfnamefont {M.}~\bibnamefont
  {Avesani}}, \bibinfo {author} {\bibfnamefont {L.}~\bibnamefont {Calderaro}},
  \bibinfo {author} {\bibfnamefont {M.}~\bibnamefont {Schiavon}}, \bibinfo
  {author} {\bibfnamefont {A.}~\bibnamefont {Stanco}}, \bibinfo {author}
  {\bibfnamefont {C.}~\bibnamefont {Agnesi}}, \bibinfo {author} {\bibfnamefont
  {A.}~\bibnamefont {Santamato}}, \bibinfo {author} {\bibfnamefont
  {M.}~\bibnamefont {Zahidy}}, \bibinfo {author} {\bibfnamefont
  {A.}~\bibnamefont {Scriminich}}, \bibinfo {author} {\bibfnamefont
  {G.}~\bibnamefont {Foletto}}, \bibinfo {author} {\bibfnamefont
  {G.}~\bibnamefont {Contestabile}}, \bibinfo {author} {\bibfnamefont
  {M.}~\bibnamefont {Chiesa}}, \bibinfo {author} {\bibfnamefont
  {D.}~\bibnamefont {Rotta}}, \bibinfo {author} {\bibfnamefont
  {M.}~\bibnamefont {Artiglia}}, \bibinfo {author} {\bibfnamefont
  {A.}~\bibnamefont {Montanaro}}, \bibinfo {author} {\bibfnamefont
  {M.}~\bibnamefont {Romagnoli}}, \bibinfo {author} {\bibfnamefont
  {V.}~\bibnamefont {Sorianello}}, \bibinfo {author} {\bibfnamefont
  {F.}~\bibnamefont {Vedovato}}, \bibinfo {author} {\bibfnamefont
  {G.}~\bibnamefont {Vallone}}, \ and\ \bibinfo {author} {\bibfnamefont
  {P.}~\bibnamefont {Villoresi}},\ }\href {\doibase 10.1038/s41534-021-00421-2}
  {\bibfield  {journal} {\bibinfo  {journal} {npj Quantum Inf}\ }\textbf
  {\bibinfo {volume} {7}},\ \bibinfo {pages} {93} (\bibinfo {year}
  {2021})}\BibitemShut {NoStop}%
\bibitem [{\citenamefont {{Yura}}(1973)}]{Yura:1973}%
  \BibitemOpen
  \bibfield  {author} {\bibinfo {author} {\bibfnamefont {H.~T.}\ \bibnamefont
  {{Yura}}},\ }\href
  {https://www.osapublishing.org/josa/abstract.cfm?uri=josa-63-5-567}
  {\bibfield  {journal} {\bibinfo  {journal} {J. Opt. Soc. Am.}\ }\textbf
  {\bibinfo {volume} {63}},\ \bibinfo {pages} {567} (\bibinfo {year}
  {1973})}\BibitemShut {NoStop}%
\bibitem [{\citenamefont {Vasylyev}\ \emph {et~al.}(2012)\citenamefont
  {Vasylyev}, \citenamefont {Semenov},\ and\ \citenamefont
  {Vogel}}]{Vasylyev:BeamWandering}%
  \BibitemOpen
  \bibfield  {author} {\bibinfo {author} {\bibfnamefont {D.~Y.}\ \bibnamefont
  {Vasylyev}}, \bibinfo {author} {\bibfnamefont {A.~A.}\ \bibnamefont
  {Semenov}}, \ and\ \bibinfo {author} {\bibfnamefont {W.}~\bibnamefont
  {Vogel}},\ }\href {\doibase 10.1103/PhysRevLett.108.220501} {\bibfield
  {journal} {\bibinfo  {journal} {Phys. Rev. Lett.}\ }\textbf {\bibinfo
  {volume} {108}},\ \bibinfo {pages} {220501} (\bibinfo {year}
  {2012})}\BibitemShut {NoStop}%
\bibitem [{\citenamefont {Arfken}\ \emph {et~al.}(2013)\citenamefont {Arfken},
  \citenamefont {Weber},\ and\ \citenamefont {Harris}}]{Arfken}%
  \BibitemOpen
  \bibfield  {author} {\bibinfo {author} {\bibfnamefont {G.~B.}\ \bibnamefont
  {Arfken}}, \bibinfo {author} {\bibfnamefont {H.~J.}\ \bibnamefont {Weber}}, \
  and\ \bibinfo {author} {\bibfnamefont {F.~E.}\ \bibnamefont {Harris}},\
  }\href@noop {} {\emph {\bibinfo {title} {Mathematical Methods for
  Physicists}}},\ \bibinfo {edition} {7th}\ ed.\ (\bibinfo  {publisher}
  {Waltham, MA, Elsevier},\ \bibinfo {year} {2013})\BibitemShut {NoStop}%
\bibitem [{\citenamefont {Er-long}\ \emph {et~al.}(2005)\citenamefont
  {Er-long}, \citenamefont {Zheng-fu}, \citenamefont {Shun-sheng},
  \citenamefont {Tao}, \citenamefont {Da-sheng},\ and\ \citenamefont
  {Guang-can}}]{Erlong:BackGNoise}%
  \BibitemOpen
  \bibfield  {author} {\bibinfo {author} {\bibfnamefont {M.}~\bibnamefont
  {Er-long}}, \bibinfo {author} {\bibfnamefont {H.}~\bibnamefont {Zheng-fu}},
  \bibinfo {author} {\bibfnamefont {G.}~\bibnamefont {Shun-sheng}}, \bibinfo
  {author} {\bibfnamefont {Z.}~\bibnamefont {Tao}}, \bibinfo {author}
  {\bibfnamefont {D.}~\bibnamefont {Da-sheng}}, \ and\ \bibinfo {author}
  {\bibfnamefont {G.}~\bibnamefont {Guang-can}},\ }\href {\doibase
  10.1088/1367-2630/7/1/215} {\bibfield  {journal} {\bibinfo  {journal} {New
  Journal of Physics}\ }\textbf {\bibinfo {volume} {7}},\ \bibinfo {pages}
  {215} (\bibinfo {year} {2005})}\BibitemShut {NoStop}%
\bibitem [{\citenamefont {Liorni}\ \emph {et~al.}(2019)\citenamefont {Liorni},
  \citenamefont {Kampermann},\ and\ \citenamefont
  {Bru{\ss}}}]{Liorni:SatQKD2019}%
  \BibitemOpen
  \bibfield  {author} {\bibinfo {author} {\bibfnamefont {C.}~\bibnamefont
  {Liorni}}, \bibinfo {author} {\bibfnamefont {H.}~\bibnamefont {Kampermann}},
  \ and\ \bibinfo {author} {\bibfnamefont {D.}~\bibnamefont {Bru{\ss}}},\
  }\href {\doibase 10.1088/1367-2630/ab41a2} {\bibfield  {journal} {\bibinfo
  {journal} {New Journal of Physics}\ }\textbf {\bibinfo {volume} {21}},\
  \bibinfo {pages} {093055} (\bibinfo {year} {2019})}\BibitemShut {NoStop}%
\bibitem [{\citenamefont {Papanastasiou}\ \emph {et~al.}(2017)\citenamefont
  {Papanastasiou}, \citenamefont {Ottaviani},\ and\ \citenamefont
  {Pirandola}}]{Papanastasiou:CVMDI}%
  \BibitemOpen
  \bibfield  {author} {\bibinfo {author} {\bibfnamefont {P.}~\bibnamefont
  {Papanastasiou}}, \bibinfo {author} {\bibfnamefont {C.}~\bibnamefont
  {Ottaviani}}, \ and\ \bibinfo {author} {\bibfnamefont {S.}~\bibnamefont
  {Pirandola}},\ }\href {\doibase 10.1103/PhysRevA.96.042332} {\bibfield
  {journal} {\bibinfo  {journal} {Phys. Rev. A}\ }\textbf {\bibinfo {volume}
  {96}},\ \bibinfo {pages} {042332} (\bibinfo {year} {2017})}\BibitemShut
  {NoStop}%
\bibitem [{\citenamefont {Papanastasiou}\ \emph {et~al.}(2022)\citenamefont
  {Papanastasiou}, \citenamefont {Mountogiannakis},\ and\ \citenamefont
  {Pirandola}}]{Papanastasiou:2022}%
  \BibitemOpen
  \bibfield  {author} {\bibinfo {author} {\bibfnamefont {P.}~\bibnamefont
  {Papanastasiou}}, \bibinfo {author} {\bibfnamefont {A.~G.}\ \bibnamefont
  {Mountogiannakis}}, \ and\ \bibinfo {author} {\bibfnamefont {S.}~\bibnamefont
  {Pirandola}},\ }\href@noop {} {\bibfield  {journal} {\bibinfo  {journal} {To
  appear}\ } (\bibinfo {year} {2022})}\BibitemShut {NoStop}%
\bibitem [{\citenamefont {Andrews}\ \emph {et~al.}(2000)\citenamefont
  {Andrews}, \citenamefont {Phillips},\ and\ \citenamefont
  {Young}}]{Andrews:2000}%
  \BibitemOpen
  \bibfield  {author} {\bibinfo {author} {\bibfnamefont {L.~C.}\ \bibnamefont
  {Andrews}}, \bibinfo {author} {\bibfnamefont {R.~L.}\ \bibnamefont
  {Phillips}}, \ and\ \bibinfo {author} {\bibfnamefont {C.~Y.}\ \bibnamefont
  {Young}},\ }\href {\doibase 10.1117/1.1327839} {\bibfield  {journal}
  {\bibinfo  {journal} {Optical Engineering}\ }\textbf {\bibinfo {volume}
  {39}},\ \bibinfo {pages} {3272 } (\bibinfo {year} {2000})}\BibitemShut
  {NoStop}%
\end{thebibliography}%
%%%%%%%%%%%%%%%%%%%%%%%%%

\end{document}